\documentclass[preprint2]{aastex62}

\newcommand{\speed}[1]{#1 km~s${}^{-1}$}
\newcommand{\accel}[1]{#1 km~s${}^{-2}$}
\newcommand{\nfig}[1]{Figure~\ref{#1}}

\newcommand{\tbl}[1]{Table~\ref{#1}}

\received{January 1, 2018}
\revised{January 7, 2018}
\accepted{\today}
\submitjournal{ApJ}

\shorttitle{Shock Waves Driven by Coronal Jets}
\shortauthors{Shen et al.}

\begin{document}

\title{Homologous large-amplitude Nonlinear fast-mode Magnetosonic Waves Driven by Recurrent Coronal Jets}

\correspondingauthor{Yuandeng Shen}
\email{ydshen@ynao.ac.cn}

\author[0000-0001-9493-4418]{Yuandeng Shen}
\affiliation{Yunnan Observatories, Chinese Academy of Sciences,  Kunming, 650216, China}
\affiliation{State Key Laboratory of Space Weather, Chinese Academy of Sciences, Beijing 100190, China}
\affiliation{CAS Key Laboratory of Solar Activity, National Astronomical Observatories, Beijing 100012, China}
\affiliation{Center for Astronomical Mega-Science, Chinese Academy of Sciences, Beijing, 100012, China}
\author[0000-0002-7694-2454]{Yu Liu}
\affiliation{Yunnan Observatories, Chinese Academy of Sciences,  Kunming, 650216, China}
\affiliation{Center for Astronomical Mega-Science, Chinese Academy of Sciences, Beijing, 100012, China}
\author[0000-0002-3483-5909]{Ying D. Liu}
\affiliation{State Key Laboratory of Space Weather, Chinese Academy of Sciences, Beijing 100190, China}
\affiliation{University of Chinese Academy of Sciences, Beijing 100049, China}
\author{Jiangtao Su}
\affiliation{CAS Key Laboratory of Solar Activity, National Astronomical Observatories, Beijing 100012, China}
\affiliation{University of Chinese Academy of Sciences, Beijing 100049, China}
\author{Zehao Tang}
\affiliation{Yunnan Observatories, Chinese Academy of Sciences,  Kunming, 650216, China}
\affiliation{University of Chinese Academy of Sciences, Beijing 100049, China}
\author{Yuhu Miao}
\affiliation{Yunnan Observatories, Chinese Academy of Sciences,  Kunming, 650216, China}
\affiliation{University of Chinese Academy of Sciences, Beijing 100049, China}
 
\begin{abstract}
The detailed observational analysis of a homologous Extreme-ultraviolet (EUV) wave event is presented to study the driving mechanism and the physical property of the EUV waves, combining high resolution data taken by the {\em Solar Dynamics Observatory} and the {\em Solar TErrestrial RElations Observatory}. It is observed that four homologous EUV waves originated from the same active region AR11476 within about one hour, and the time separations between consecutive waves were of 8 -- 20 minutes. The waves showed narrow arc-shaped wavefronts and propagated in the same direction along a large-scale transequatorial loop system at a speed of \speed{648 -- 712} and a deceleration of \accel{0.985 -- 1.219}. The EUV waves were accompanied by weak flares, coronal jets, and radio type \uppercase\expandafter{\romannumeral3} bursts, in which the EUV waves were delayed with respect to the start times of the radio type \uppercase\expandafter{\romannumeral3} bursts and coronal jets about 2 -- 13 and 4 -- 9 minutes, respectively. Different to previous studies of homologous EUV waves, no coronal mass ejections were found in the present event. Based on the observational results and the close temporal the spatial relationship between the EUV waves and the coronal jets, for the first time, we propose that the observed homologous EUV waves were large-amplitude nonlinear fast-mode magnetosonic waves or shocks driven by the associated recurrent coronal jets, resemble the generation mechanism of a piston shock in a tube. In addition, it is found that the recurrent jets were tightly associated with the alternating flux cancellation and emergence in the eruption source region and radio type \uppercase\expandafter{\romannumeral3} bursts.
\end{abstract}


\section{Introduction}
Large-scale extreme-ultraviolet (EUV) waves have been the subject of extensive study for twenty years. They appear as bright arc-shaped wavefronts and can propagate hundreds of megameters under the quiet-Sun condition \citep{1998GeoRL..25.2465T}. Recent statistical studies based on high temporal and high spatial resolution observations taken by the Atmospheric Imaging Assembly \citep[AIA;][]{2012SoPh..275...17L} on board the {\em Solar Dynamics Observatory} \citep[{\em SDO};][]{2012SoPh..275....3P} indicated that EUV waves have a board speed distribution of \speed{200 -- 1500}, and at an average speed of \speed{664} \citep{2013ApJ...776...58N}. These results are much higher than those obtained in statistical studies based on lower resolution observations taken by the Extreme Ultraviolet Telescope \citep[EIT;][]{1995SoPh..162..291D} on board the {\em Solar and Heliospheric Observatory ({\em SOHO})} and the the Extreme Ultraviolet Imager \citep[EUVI;][]{2004SPIE.5171..111W} on board the {\em Solar Terrestrial Relations Observatory} \citep[{\em STEREO};][]{2008SSRv..136....5K}. For example, based on {\em SOHO}/EIT data, \cite{2009ApJS..183..225T} found that the speed of EUV waves ranges from 50 to \speed{700}, and the typical speed is about \speed{200 -- 400}. The statistical results based on {\em STEREO}/EUVI revealed that the average speeds of EUV waves are about \speed{300} \citep[e.g.,][]{2011A&A...532A.151W,2014SoPh..289.4563M,2014SoPh..289.1257N}. It seems that the speeds of EUV waves are significantly affected by the temporal and spatial resolution of the observations. Recently, based on the {\em SDO}/AIA high resolution data, \cite{2017SoPh..292..185L} found that the average initial speed of EUV waves is about \speed{360}, which is much more consistent with the earlier results derived from {\em SOHO}/EIT and {\em STEREO}/EUVI data. If this is true, the cadence and spatial resolution do not affect the measuring results about the speed of EUV waves. \cite{2017SoPh..292..185L} found that about 45\% solar eruptive events are accompanied by EUV waves, and they are associated with a variety of solar eruptive phenomena such as flares, coronal mass ejections (CMEs), Type \uppercase\expandafter{\romannumeral2} radio bursts, and EUV dimmings \citep{2011SSRv..158..365G,2005ApJ...625L..67V,2006AA...448..739V}. Study on EUV waves can provide important clues for diagnosing other solar eruptive phenomena. In addition, since EUV waves can propagate a large distance and with a large angular extent, they are thought to be an important agent to trigger transverse and longitudinal oscillations of remote large-scale filaments \citep{2004ApJ...608.1124O,2012ApJ...754....7S,2014ApJ...786..151S,2014ApJ...795..130S,2017ApJ...851..101S,2016SoPh..291.3303P}, coronal loops \citep[e.g.,][]{1999SoPh..190..467W,2012ApJ...753...52L}, coronal cavities \citep[e.g.,][]{2012ApJ...753...52L,2018arXiv180501088Z}, and sympathetic solar events at different locations \citep{2014ApJ...786..151S}. In addition, the observational results of EUV waves can be used to diagnose the coronal magnetic fields \citep[e.g.,][]{1999ESASP.446..477M,2005A&A...435.1123W,2011ApJ...730..122W,2017AA...603A.101L}.

Earlier studies found that the propagation of EUV waves appear to stop at coronal hole boundaries and avoid active regions \citep{1999ApJ...517L.151T}. It is unclear that the observed EUV waves are real magnetohydrodynamics (MHD) waves or not. In recent years, many studies based on high resolution data revealed that wave phenomena such as reflection \citep[e.g.,][]{2008ApJ...680L..81L,2009ApJ...691L.123G,2012ApJ...752L..23S,2012ApJ...746...13L,2012ApJ...756..143O,2013ApJ...775...39Y}, transmission \citep[e.g.,][]{2012ApJ...753...52L,2012ApJ...756..143O,2013ApJ...773L..33S}, refraction \citep[e.g.,][]{2012ApJ...754....7S,2013ApJ...773L..33S}, and mode conversion \citep[e.g.,][]{2016SoPh..291.3195C,2017ApJ...834L..15Z} are frequently observed when EUV waves interact with other magnetic structures such as coronal holes, quiet-Sun bright points, and active regions. Especially, \cite{2013ApJ...773L..33S} simultaneously observed these wave phenomena in a single EUV wave when it passed through two remote active regions. These observations confirmed the scenario that EUV waves are fast-mode MHD waves in nature. In addition, as predicted by theory \citep{1968SoPh....4...30U}, the on-disk EUV waves are in fact the intersection line of a dome-shaped shock in the corona. Therefore, EUV waves should cause wave signals at different atmosphere heights, and their shapes and speeds should be roughly similar. This has been confirmed by the observational study performed by \cite{2012ApJ...752L..23S}, in which the authors observed quasi-cospatial wavefronts from the upper photosphere to the low corona, and the wave speeds at different atmosphere layers are similar during the initial stage. Strictly speaking, the speed of an EUV wave should not be exactly the same, because the local fast-mode speed increases with height in the quiet corona. This is consistent with observations of limb events where the wavefronts become increasingly tilted towards the solar surface \citep[e.g.,][]{2003SoPh..212..121H,2009ApJ...700L.182P}. However, in most observations the speed differences at different heights are too small to discern. Recently, \cite{2016SoPh..291...89V} and \cite{2018MNRAS.474..770K} found in their simulation that only powerful EUV waves with strong lateral expansion can cause chromospheric response, consistent with the observational results and predictions in \cite{2012ApJ...752L..23S}. Furthermore, \cite{2012ApJ...752L..23S} also confirmed the co-existing fast and slow wave components as predicted in numerical studies \citep{2002ApJ...572L..99C}, in which the fast wave component is a fast-mode MHD wave, while the slow one is possibly formed due to successive stretching of the magnetic field lines. These observational results are in good agreement with the theoretical prediction that the propagation speed of the slow wave-like component is  about one third of the fast-mode MHD wave \citep[see also;][]{2012ApJ...752L..23S,2013AA...553A.109K,2015ApJ...809..151Z}. However, it should be noted that some authors also proposed that the slow wave component can be regraded as slow-mode magnetosonic waves \citep[e.g.,][]{2009ApJ...700.1716W,2012SCPMA..55.1316M}. 

The driving mechanism of EUV waves is another unsolved problem. Some authors proposed that EUV waves are driven by flare pressure pulses \citep[e.g.,][]{2002AA...383.1018K,2003SoPh..212..121H,2004AA...418.1117W}, whereas others believed in that EUV waves are excited by CMEs \citep[e.g.,][]{2002ApJ...572L..99C,2006ApJ...641L.153C,2012ApJ...752L..23S,2017ApJ...851..101S,2017SoPh..292....7L,2013AA...556A.152X}. \cite{2006ApJ...641L.153C} performed a statistical survey with 14 non-CME-associated energetic flares that should possess strong pressure pulses for driving EUV waves. However, the author did not find any EUV waves in association with these flares. In addition, the author also studied an active region that hosts both CME-associated and non-CME types of flares. Their result indicated that EUV waves only appear when CMEs are present. This study indicates that EUV waves should be driven by CMEs rather than flare pressure pulses. By using stereoscopic observations taken by {\em STEREO}/EUVI and three-dimensional geometrical modeling of the CME and wave structures, \cite{2009SoPh..259...49P} found that the EUV wave occupies and affects a much bigger volume than the CME; meanwhile, they observed the actual detachment of the wave from the CME \citep{2009ApJ...700L.182P}. Therefore, the authors suggested that the observed EUV wave and the associated CME were separated in space, and the former is likely driven by the latter \citep[see also,][]{2009ApJ...703L.118K,2011ApJ...734...84L}. Observations of the detailed driving processes of EUV waves by expanding loops or cavities were presented in several studies by using high temporal and high spatial resolution data taken by {\em SDO}/AIA. \cite{2010A&A...522A.100P} reported the detachment of an EUV wave from the CME volume shortly after the impulsive cavity expansion. \cite{2011ApJ...738..160M} observed a dome-like EUV wave (shock) propagated ahead of the following semi-spherical CME bubble. \cite{2012ApJ...745L...5C} studied a limb EUV wave on 2011 June 7, and they observed the detailed separation process between the EUV wave and the associated CME in the low corona. These studies provide primarily observational evidence for supporting the scenario that EUV waves are driven by CMEs. Above studies are all large-scale energetic solar eruption events. The study of less energetic solar eruptive events is also important for determining the driving mechanism and physical nature of EUV waves. Recently, \cite{2017ApJ...851..101S} studied a small-scale EUV wave in association with a micro-flare (B1.9) in a quiet-Sun region, in which the authors identified the detailed separation process between the EUV wave and a group of expanding coronal loops that was caused by the eruption of a mini-filament. Small-scale EUV waves are also observed in association with failed filament eruption \citep{2012ApJ...753..112Z,2012AA...541A..49Z} and newly formed expanding loops through tether-cutting reconnection in micro-sigmoid structures \citep{2012ApJ...753L..29Z,2013MNRAS.431.1359Z}. Small-scale solar eruptive events have smaller size and energy scale than large-scale filament eruptions or CMEs, and they share many common characteristics with their large-scale counterparts. For example, both small- and large-scale solar eruptive phenomena are associated with the ejection of plasma, and triggered or driven by magnetic reconnection. In the past two decades, a large number of articles aimed to study the wave nature and driving mechanism, and more details about EUV wave can be found in recent review papers \citep[][and references therein]{2014SoPh..289.3233L,2015LRSP...12....3W}.

Homologous eruptive events are frequently observed in the solar atmosphere from small- to large-scale at different regions such as quiet-Sun regions, active regions, and coronal holes. They often occur in the same source region within a short time interval from a few minutes to several hours. So far, different kinds of homologous solar events have been documented in the literature. For example, the homologous flares \citep{2001JGR...10625227S,2004ApJ...612..546S,2017ApJ...850....8J,2017ApJ...851...30X,2018ApJ...852L..10R,2014ApJ...791...84W}, recurrent coronal jets \citep{2008AA...478..907C,2010ApJ...714.1762P,2012ApJ...760..101W,2015ApJ...815...71C,2015ApJ...801...83C,2016ApJ...833..150L,2016ApJ...822L..23P,2016SoPh..291..859Z,2017ApJ...845...94T,2017ApJ...851...67S}, successive filament eruptions \citep{2008ApJ...680..740D, 2011RAA....11..594S,2012ApJ...750...12S,2012ApJ...745....9Y,2016ApJ...827L..12W,2012AJ....143...56Y,2018ApJ...856...79Y}, homologous cyclones \citep{2014ApJ...782L..15Y}, and coronal mass ejections \citep[CME;][]{2002ApJ...566L.117Z,2012ApJ...750...12S,2013ApJ...763L..43W,2013ApJ...769...45L,2014ApJ...788L..28L,2017ApJ...845...59V,2017SoPh..292...64L}. Many studies indicated that homologous eruptive events often have the same trigger and eruption mechanisms, and most of them are found to be in association with magnetic flux emergence and cancellation in the photosphere. Due to the similar background conditions, trigger mechanisms, and eruption processes, studies of homologous solar eruptive events are useful for diagnosing the correlation between different physical parameters.

Although a large number of EUV wave events reported in previous studies, the observations of homologous EUV waves are very scarce. To our knowledge, so far only two homologous EUV wave events were studied in detail \citep{2011ApJ...727L..43K,2012ApJ...747...67Z}. Using observations taken by {\em STEREO}/EUVI, \cite{2011ApJ...727L..43K} reported the first homologous EUV wave event on 2010 April 28 to 29. The authors identified four EUV waves from the same source active region within eight hours. These waves propagated along the same path at a constant speed of \speed{220 -- 340}, and each wave was accompanied by a weak flare and a faint CME. The authors found that the magnetosonic Mach numbers and speeds of the four waves are distinctly correlated, and therefore proposed that these waves are nonlinear fast-mode magnetosonic waves in nature. \cite{2012ApJ...747...67Z} studied the second homologous EUV wave event observed by the {\em SDO} on 2010 November 11. The authors detected four EUV waves originated from active region AR11124 within three hours. Similar to the event studied by \cite{2011ApJ...727L..43K}, these waves also emanated from the same source region and propagated in the same direction, and all of the four EUV waves were accompanied by weak flares and faint CMEs. The propagation speeds of the waves were also at a constant value of \speed{280 -- 500}, and hence they are interpreted as fast-mode waves. By analyzing the eruptive events in the source region and the variations of the photospheric magnetic fluxes, the authors further suggested that the observed homologous waves were tightly associated with surges and continuous emergence and cancellation of magnetic fluxes. In addition, \cite{2008ApJ...684L..45N} reported three successive Moreton waves generated by a single solar flare on 2005 August 3, and they proposed that these waves are caused by the successive filament eruptions (also three times) in the source region. While there is no direct proof of homologous EUV waves for the Moreton waves reported by \cite{2008ApJ...684L..45N}, it is however very likely, since all Moreton waves where EUV data was available were associated with EUV waves \citep[see][]{2010AdSpR..45..527W}.  

So far, solar physicists have widely accepted that EUV waves are fast-mode magnetosonic waves in nature and are driven by CMEs. Generally speaking, any disturbance can launch waves in the solar atmosphere. Therefore, other kinds of solar eruptive events are also potential drivers for EUV waves. In this paper, we present an interesting homologous EUV wave event on 2012 May 14 from NOAA active region AR11476. Four EUV waves are observed within one hour. The propagation of the waves are along the same transequatorial loop system, and each of them was associated with a weak flare and a coronal jet in the eruption source region, and a radio type \uppercase\expandafter{\romannumeral3} burst. Different to previous reported homologous EUV waves, the wavefronts of the EUV waves in the present study are very narrow, and without CME association. The time separations between consecutive EUV waves and the shapes of their wavefronts are much different to those of quasi-periodic fast propagating (QFP) EUV waves that often have a funnel-like shape and propagate along coronal loops \citep[e.g.,][]{2011ApJ...736L..13L,2012ApJ...753...52L,2012ApJ...753...53S,2013SoPh..288..585S,2018ApJ...853....1S,2018MNRAS.477L...6S}. Therefore, we can not consider the observed homologous EUV waves as QFP waves. Instruments and observations are introduced in Section 2. Main observational results are described in Section 3. Discussion and conclusion are presented in the last section.

\begin{figure*}[thbp]
\epsscale{1}
\figurenum{1}
\plotone{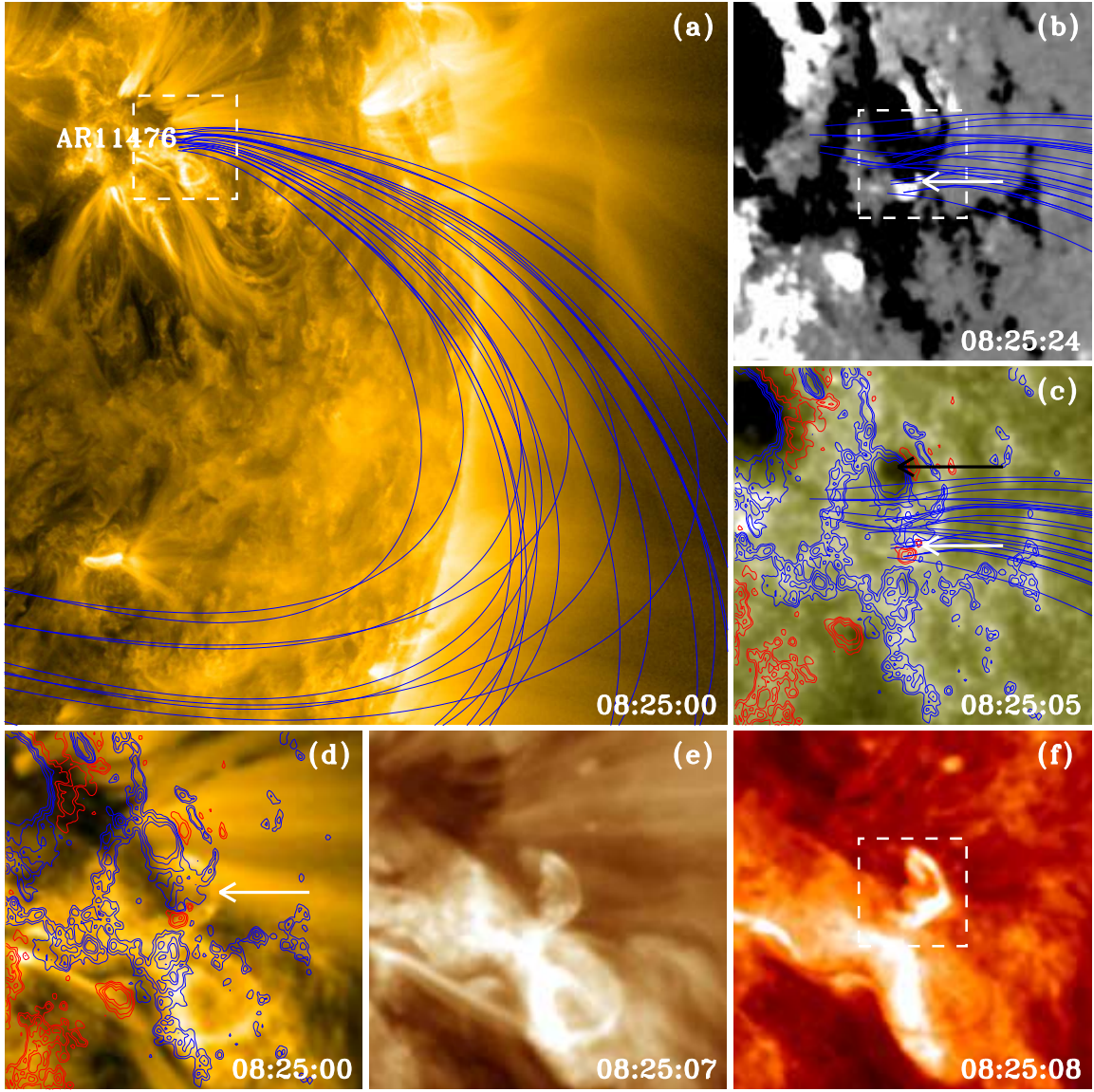}
\caption{An overview of the pre-eruption magnetic condition. Panel (a) is an AIA 171 \AA\ image overlaid with the extrapolated magnetic field lines (blue curves), The close-up view of the eruption source region is shown in the rest of panels, in which panel (b) is an HMI LOS magnetogram, and panels (c) -- (f) are the AIA 1600, 171, 193, and 304 \AA\ images, respectively. In panel (b), the white (black) patches represent positive (negative) magnetic polarities. The positive (negative) magnetic fields at 08:25:24 UT are also overlaid in panels (c) and (d) as red (blue) contours. The arrow in panel (d) points to the small loop structure, and the field lines in panel (a) are also overlaid in panels (b) and (c). The field-of-view (FOV) of panel (a) and the other panels are 700\arcsec $\times$ 700\arcsec and 100\arcsec $\times$ 100\arcsec, respectively. 
\label{fig1}}
\end{figure*}

\section{Instruments and Observations}
The present homologous EUV wave event was simultaneously observed by the {\em SDO} and {\em STEREO} Ahead ({\em STEREO}-A) from two different view angles. On 2012 May 14, the separation angle between the two satellite was about $115^{\circ}$. The {\em SDO}/AIA provides full-disk images of chromosphere and corona in seven EUV and three UV-visible channels, spanning a temperature range from approximately $2 \times 10^{4}$ K to in excess of $20 \times 10^{6}$ K. The pixel size of the AIA images is $0\arcsec.6$, and the cadences of the EUV and UV-visible images are 12 and 24 seconds, respectively. In the meantime, the Helioseismic and Magnetic Imager \citep[HMI;][]{2012SoPh..275..327S} on board {\em SDO} provides full-disk line-of-sight (LOS) magnetograms of the photosphere, whose cadence and measuring precision are 45 seconds and 10 Gauss, respectively. {\em STEREO}-A also takes full-disk EUV images in four channels. We use the 304 and 195 \AA\ images in the present study. The pixel size of the {\em STEREO} EUV images is $1\arcsec.58$, while the cadences of the 195 \AA\ and 304 \AA\ images are 5 and10 minutes, respectively. In addition, the white-light images taken by {\em SOHO}/LASCO C2 and {\em STEREO} Cor1 ahead coronagraphs are also used. The soft X-ray fluxes recorded by the {\em Geostationary Operational Environmental Satellite} ({\em GOES}), and the hard X-ray fluxes recorded by the {\sl Reuven Ramaty High Energy Solar Spectroscopic Imager (RHESSI)} \citep{2002SoPh..210....3L} are also used to analyze the flares in the eruption source region. The radio spectrometer provided by the ``Orbita'' Callisto station, which is located near Almaty, is used to diagnose the magnetic reconnection processes that cause the homologous jets. The radio telescope operates between 45 and 870 MHz having a frequency resolution of 62.5 KHz and a temporal resolution of 0.25 s \citep{2014SunGe...9...71Z}. All images used in this paper are differentially rotated to the reference time of 09:18:00 UT on May 14, and the solar north is up, west to the right.

\section{Observational Results}
On 2012 May 14, four recurrent coronal jets occurred in NOAA active region AR11476 (N09, W47) within about one hour from 08:45:00 to 09:45:00 UT. During the ejection time interval, besides the flares in the eruption source region, it is interesting that each jet was accompanied by a bright narrow EUV wave in the AIA 171 \AA\ observations. The waves were ahead of the jets and propagated along the same trajectory. By checking the {\em GOES} soft X-ray fluxes, it is found that each jet/wave was associated with a weak micro-flare. The {\em GOES} class of the flares are of B4.9, B4.4, B5.2, and C2.5, and their corresponding start times are 08:44:20, 09:00:15, 09:09:18, and 09:35:05 UT, respectively. The white-light coronagraph images recorded by {\em SOHO}/LASCO and {\em STEREO}/Cor1 revealed that only the last jet/wave was accompanied by a small jet-like CME. This result is inconsistent with previous findings where homologous EUV waves were all accompanied by faint CMEs \citep{2011ApJ...727L..43K,2012ApJ...747...67Z}.

The pre-eruption magnetic condition is displayed in \nfig{fig1} by using multi-wavelength AIA and HMI  observations. Panel (a) is an AIA 171 \AA\ image overlaid with the extrapolated magnetic field lines (blue curves) based on the potential field source surface (PFSS) model \citep{2003SoPh..212..165S}. It can be seen that the location of the eruption source region (AR11476) was in the northern hemisphere. The PFSS extrapolation revealed a transequatorial loop system that connected AR11476 in the northern hemisphere and AR11480 (not show here) in the southern hemisphere. This loop system served as the propagation path of the observed jets and waves. The details of the eruption source region, which indicates with a dashed white box in \nfig{fig1} (a), is shown in \nfig{fig1} (b) -- (f). One can see that AR11476 was primarily composed of a main and a small satellite sunspot (see \nfig{fig1} (c)), and both of them were of negative magnetic polarity. The footpoint of the transequatorial loop system was rooted in a negative magnetic region between the two sunspots. In addition, there was a small positive magnetic polarity that can be identified close to the small satellite sunspot (see the black and white arrows panels \nfig{fig1} (b) and (c)). In EUV observations (panels (d) -- (f) of \nfig{fig1}), there was a small bright loop-like structure that connected the small positive magnetic polarity and the satellite sunspot (see the arrow in \nfig{fig1} (d) and the box region in \nfig{fig1} (f)). Such a magnetic morphology is in favor of the occurrence of coronal jets \citep[e.g.,][]{1994ApJ...431L..51S,2007Sci...318.1591S,2011ApJ...735L..43S}.

\begin{figure*}[thbp]
\epsscale{1}
\figurenum{2}
\plotone{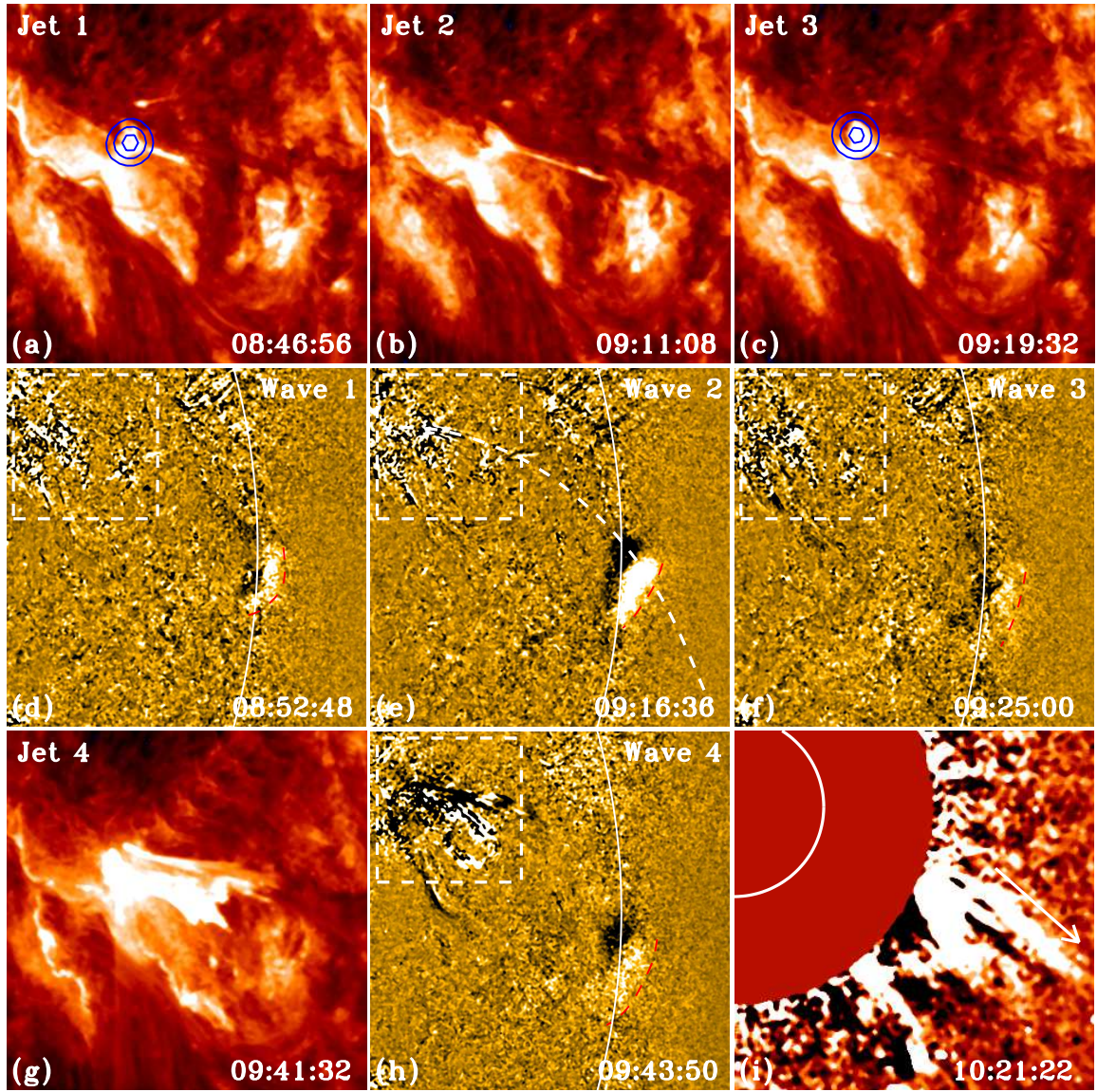}
\caption{Panels (a) -- (c) and (d) show the four recurrent jets with the AIA 304 \AA\ images, while panel (d) -- (f), and (h) show the homologous EUV waves with the AIA 171 \AA\ running difference images. Panel (i) is a LASCO C2 running difference image shows the jet-like CME. The white dashed boxes overlaid in the AIA 171 \AA\ images indicate the FOV of the AIA 304 \AA\ images, and the white circle indicates the size of the Sun. The blue contours in panels (a) and (c) indicate the {\em RHESSI} X-ray source, while the dashed red curves in the 171 \AA\ images indicate the bright wavefronts. The white curves in panels (d) -- (f) and (h) indicate the solar limb, while the white dashed curve in panel (e) indicates the path along which time-distance stack plots are obtained. The FOVs of the AIA 304 \AA\, 171 \AA\, and the LASCO images are 200\arcsec $\times$ 200\arcsec, 500\arcsec $\times$ 500\arcsec, and 3800\arcsec $\times$ 3800\arcsec, respectively. An animation is available for this figure in the online journal.
\label{fig2}}
\end{figure*}

The four coronal jets are displayed in \nfig{fig2} (a) -- (c) and (g) in the AIA 304 \AA\  direct images, respectively. The first three jets started at about 08:45:40 UT, 09:05:08 UT, and 09:14:44 UT, respectively. They showed similar structural characteristics in the EUV observations, i.e., a bright collimated plasma structure and an inverse Y-shaped bright patch at the jet-base (see \nfig{fig2} (a)). This suggests that these jets were possibly resulting from the magnetic reconnection between closed and open magnetic field lines, in agreement with the typical jet model \citep[e.g.,][]{2007Sci...318.1591S,2011ApJ...735L..43S,2012RAA....12..573C}. The eruption of the fourth jet started at about 09:38:56 UT, which is the most energetic and spectacular one among the four jets, and it directly ejected into the FOV of the LASCO C2 and formed the observed jet-like CME. It should be pointed out that there was a dark loop-like feature adjacent to the ejecting jet plasma on the norther side in the AIA 304 \AA\ images. It was not the so-called cool component as observed in some coronal jets \citep[e.g.,][]{2012ApJ...745..164S,2017ApJ...851...67S}, where the jets' cool components were observed as the erupting materials of mini-filaments. Here, we think that the dark loop-like feature was possibly cool loops as reveled by the PFSS extrapolation. By using the {\em RHESSI} observations and the reconstruction software available in the SolarSoftware (SSW) package, we reconstructed the X-ray sources at the jet base with the clean algorithm. For the first and the third jets, we successfully reconstructed the X-ray sources in the energy band of 6 -- 12 keV, and the results are overlaid as blue contours in \nfig{fig2} (a) and (c). It can be seen that the X-ray sources are well overlapped with the bright inverse Y-shaped bright structures that represent the reconnection sites of the jets. We failed to reconstruct the X-ray source for the second jet, since the flare associated with this jet is very weak. For the fourth jet, {\em RHESSI} did not recorded data during that time interval.

\begin{figure*}[thbp]
\epsscale{1}
\figurenum{3}
\plotone{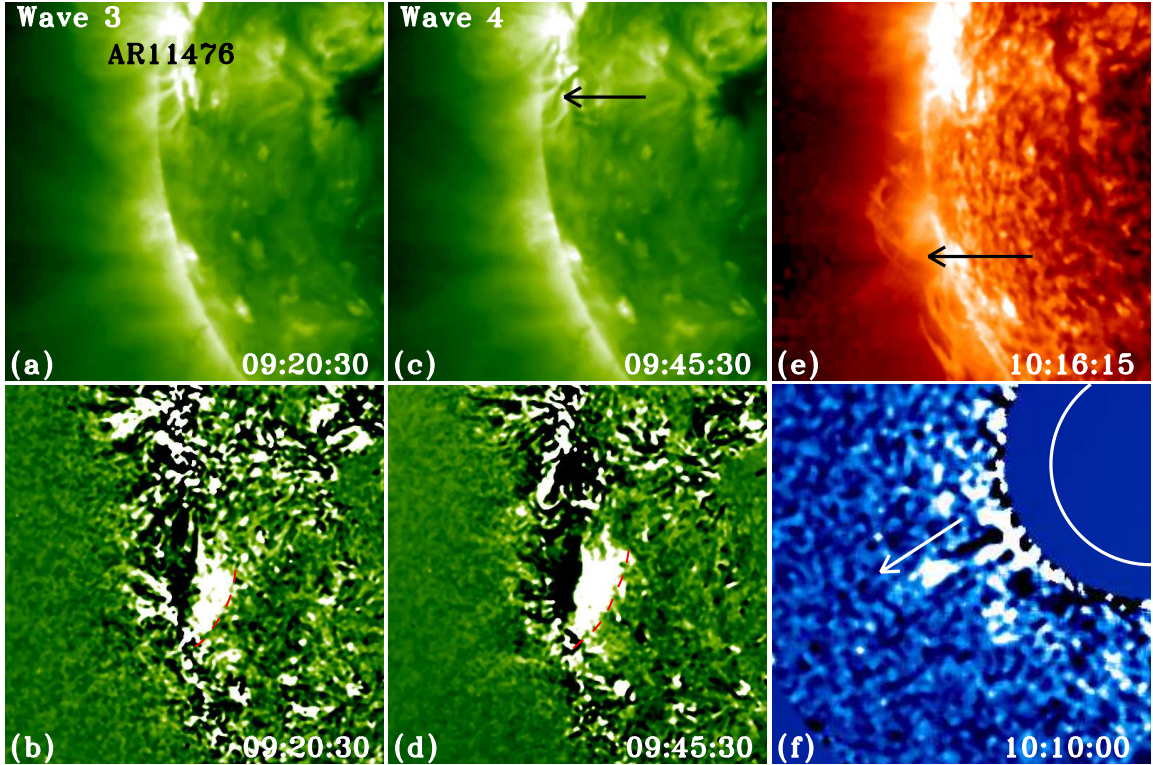}
\caption{Panel (a) and (c) are {\em STEREO}-A 195 \AA\ images, and the corresponding running difference at the same time are shown in panels (b) and (d), respectively. Panels (e) and (f) are {\em STEREO}-A 304 \AA\ and Cor1 images. The arrows in panels (c) and (e) indicate the jets, and the one in panel (f) indicates the jet-like CME. The dashed red curves in panels (b) and (d) indicate the bright wavefronts. The FOVs of the {\em STEREO} 304 and 195 \AA\ images is 850\arcsec $\times$ 850\arcsec, and that of the Cor1 images is 3800\arcsec $\times$ 3800\arcsec.
\label{fig3}}
\end{figure*}

It is interesting that the ejection of each jet was accompanied by a bright EUV wave along the transequatorial loop system as shown in \nfig{fig1} (a). The four waves are shown in \nfig{fig2} (d) -- (f) and (h) using the AIA 171 \AA\ running difference images, and the wavefront of each wave is highlighted with a dashed red curve. Here, a running difference image is obtained by subtracting the present image by the previous one in time, and moving features can be observed clearly in running difference images. As one can see that the wavefronts are very narrow relative to those observed in normal EUV waves that often have an arc-shaped structure and with a large angular extent \citep[e.g.,][]{1998GeoRL..25.2465T,2013ApJ...773L..33S,2013ApJ...776...58N}. A dark dimming region can be identified behind each EUV wave in the running difference images, but this can be artificial due to the algorithm of running difference images. Therefore, we further checked the base ratio images and the dimming regions did appear behind the EUV waves (not show here). This phenomenon is similar to normal large-scale EUV waves \citep[e.g.,][]{2012ApJ...754....7S,2013SoPh..286..509L}. In coronagraph observations, we do not find any associated CMEs for the first three coronal jets. However, the fourth jet did caused a small jet-like CME, but it just the extension of the jet structure in the FOV of the coronagraphs (see the white arrow in \nfig{fig2} (i)). More details about the EUV waves and jets, one can see the animation available in the online journal.

\begin{figure*}
\epsscale{1}
\figurenum{4}
\plotone{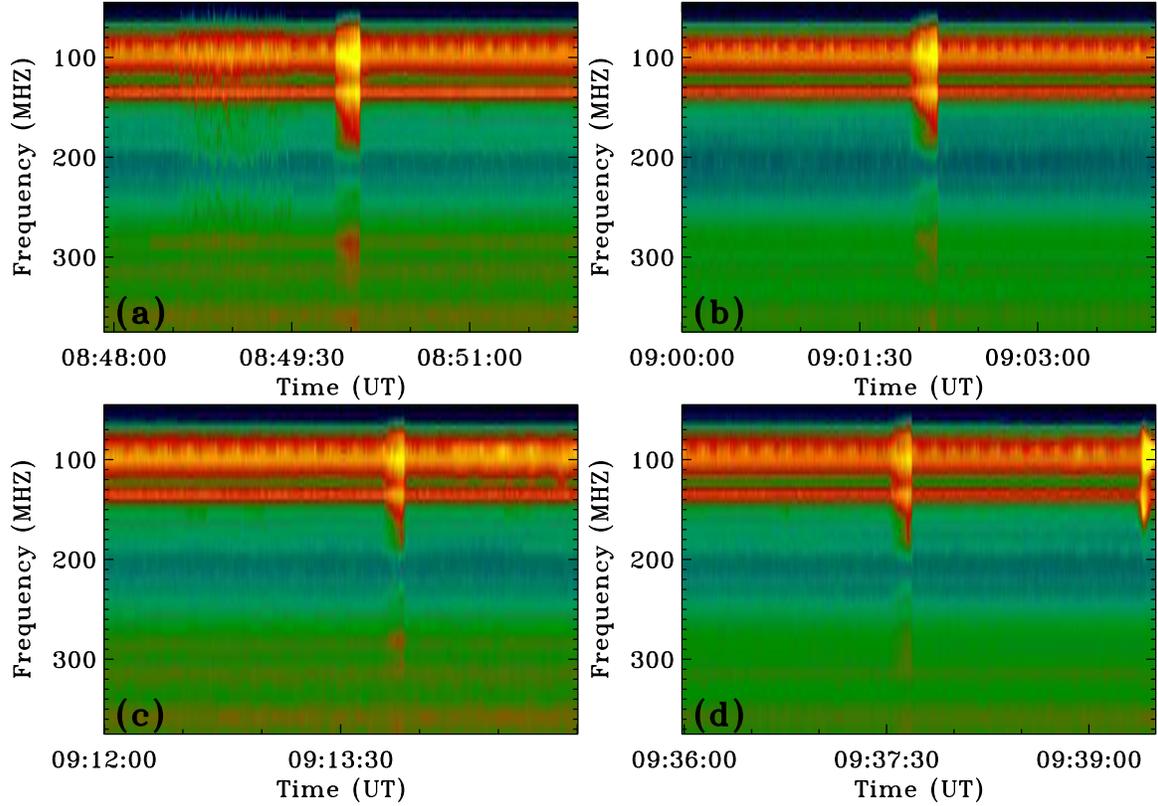}
\caption{Callisto (ALMATY) radio spectrums show the radio type \uppercase\expandafter{\romannumeral3} bursts associated with the jets. Here, we only show the frequency range from 45 to 375 MHZ. Panels (a) to (d) show the radio type \uppercase\expandafter{\romannumeral3} bursts associated with the first to fourth jets, respectively.
\label{fig4}}
\end{figure*}

The third and the fourth EUV waves were also observed by the {\em STEREO}-A from the other view angle, and the observations are displayed in \nfig{fig3}. Due to the low temporal and spatial resolution of the {\em STEREO} observations, only the largest jet (the fourth one) can be identified (see the arrow in \nfig{fig3} (c) and (e)). The wavefronts of the third and the fourth waves can be clearly observed in the 195 \AA\ running difference images (see \nfig{fig3} (b) and (d)). In addition, the jet-like CME associated with the fourth jet was also detected by {\em STEREO}/Cor1 (see \nfig{fig3} (f)). In radio observations, it is found that the four homologous jets were associated with four radio type \uppercase\expandafter{\romannumeral3} bursts as displayed in \nfig{fig4}, and the start times of the radio type \uppercase\expandafter{\romannumeral3} bursts were in the impulsive rising phases of the associated flares. Generally, the appearance of of radio type \uppercase\expandafter{\romannumeral3} bursts are caused by nonthermal electrons accelerated in reconnection processes. Therefore, the appearance of radio type \uppercase\expandafter{\romannumeral3} bursts provide evidence for the occurrence of magnetic reconnection between the bright closed loop structure and the ambient large-scale transequatorial loop system \citep[see also,][]{2017ApJ...851...67S}.

\begin{figure*}[thbp]
\epsscale{1}
\figurenum{5}
\plotone{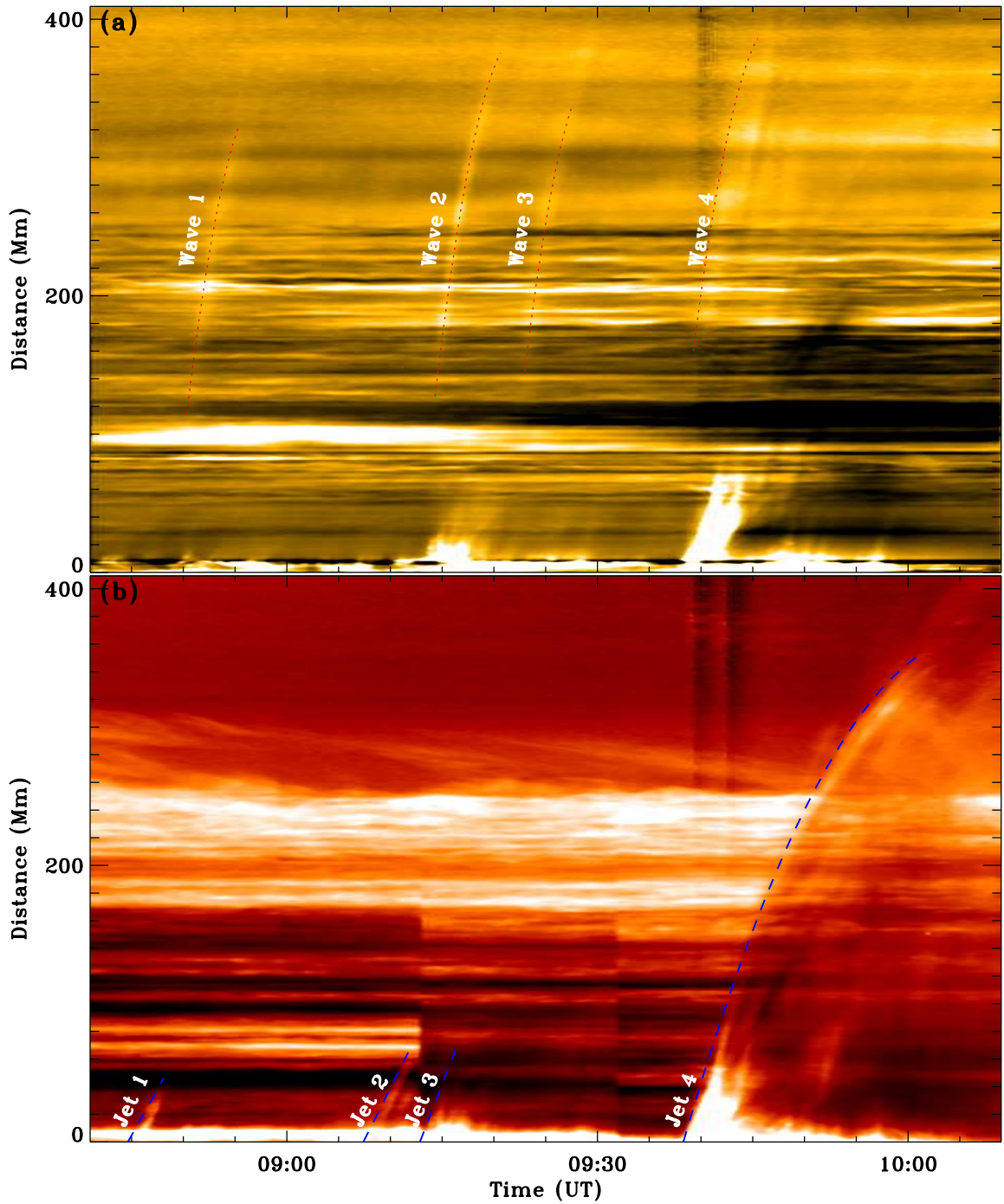}
\caption{Panels (a) is the time-distance stack plot made from AIA 171 base-ratio images along the dashed white curve as shown in \nfig{fig2} (e), in which the four EUV waves are marked with red dotted curves. Panel (b) is the time-distance made from 304 \AA\ direct images along the same path as panel (a), in which the first three jets are indicated with blue dashed lines, while the last one is indicated with a blue dashed curve.
\label{fig5}}
\end{figure*}

\begin{deluxetable*}{cccccccccc}
%\tablenum{1}
\tablecaption{Parameters of the associated flares, jets, radio type \uppercase\expandafter{\romannumeral3} bursts, and EUV waves\label{tbl1}}
\tablewidth{0pt}
\tablehead{
\colhead{Items} & \colhead{Flare$_{\rm C}$}  & \colhead{Flare$_{\rm S}$} & \colhead{Radio$_{\rm S}$}  & \colhead{Jet$_{\rm T}$} & \colhead{Jet$_{\rm S}$}&\colhead{Wave$_{\rm T}$} & \colhead{Wave$_{\rm S}$} & \colhead{Wave$_{\rm D}$} & \colhead{Wave$_{\rm IS}$} \\
\colhead{ } & \colhead{{\em GOES}}  & \colhead{(UT)} & \colhead{(UT)} & \colhead{(UT)} & \colhead{(km s$^{-1}$)} &\colhead{(UT)} & \colhead{(km s$^{-1}$)} & \colhead{(km s$^{-2}$)} & \colhead{(km s$^{-1}$)}
}
%\decimalcolnumbers
\startdata
Wave 1 &B4.9 &08:44:20 &08:49:52 &08:45:40 & 226 &08:51:24 &694 &1.175 &921 \\
Wave 2 &B4.4 &09:00:15 &09:01:55 &09:05:08 & 251 &09:14:48 &664  &1.101 &977 \\
Wave 3 &B5.2 &09:09:18 &09:13:45 &09:14:44 & 325 &09:23:00 &712  &1.219 &950 \\
Wave 4 &C2.5 &09:35:05 &09:37:52 &09:38:56 & 347 &09:42:38 &648  &0.985 &\nodata \\
\enddata
\tablecomments{Flare$_{\rm C}$ and Flare$_{\rm S}$ represent the {\em GOES} class and the start time of the flares, respectively. Radio$_{\rm S}$, Jet$_{\rm T}$ and Wave$_{\rm T}$ are the start times of the radio type \uppercase\expandafter{\romannumeral3} bursts, jets, and waves; Jet$_{\rm S}$, Wave$_{\rm S}$, and   Wave$_{\rm IS}$ are the speed of the jets, the mean speed of the EUV waves, and the initial speed of the EUV waves, respectively. Wave$_{\rm D}$ is the mean deceleration of the EUV waves.}
\end{deluxetable*}

The kinematics of the homologous waves and the jets are studied by using time-distance stack plots of intensities taken along the path as shown by the dashed white curve (in the plane-of-sky) in \nfig{fig2} (e). \nfig{fig5} (a) and (b) are time-distance stack plots made from the AIA 171 \AA\ base-ratio and 304 \AA\ direct images, respectively. To obtain a time-distance stack plot, we first obtained the one-dimensional intensity profiles along a specified path at different times, and then a two-dimensional time-distance stack plot was generated by stacking the obtained one-dimensional intensity profiles in time. Here, as one can see in \nfig{fig5}, the EUV waves and the jets can be best identified in the 171 and 304 \AA\ time-distance stack plots, respectively. The four homologous waves showed as four bright curved stripes in the AIA 171 \AA\ time-distance stack plot, and they appeared at a minimum distance of about 120 Mm from the eruption source region (see \nfig{fig5} (a)). The dimming signal behind each wave can be identified in the time-distance stack plot, but it is very weak. It can be seen that the four EUV waves underwent obvious deceleration during their propagation, rather than with constant speed as reported in previous homologous EUV wave events \citep{2011ApJ...727L..43K,2012ApJ...747...67Z}.

To obtain the speeds of the waves, we tracked the first three waves with the semi-automated method proposed by \cite{2011AA...531A..42L}, while the leading edge of the fourth wave was determined manually by eye since it is too ambiguous in the time-distance stack plot. The mean speed and deceleration of the waves can be obtained by fitting the measured data points with linear and quadratic functions, respectively. The fitting results indicate that the mean speeds of the EUV waves are \speed{683, 659, 702, and 638}, while their corresponding decelerations are \accel{1.157,1.084,1.200, and 0.970}, respectively. Here, the measured parameters of the waves are all in the plane-of-sky. Since the EUV waves propagated roughly along the transequatorial loop system, we can therefore determine their true speeds based on the extrapolated three-dimensional magnetic field. To do this, the tilt angles of the extrapolated loops as shown in \nfig{fig1} (a) were calculated at the first, then we used the average tilt angle of the loops as the tilt angle of the loop system. It is measured that the average tilt angle of the loop system with respect to the plane-of-sky is about $10^\circ$. According to the trigonometric function, the true speeds and decelerations of the EUV waves can be derived based on the tilted angle and the measured speeds. The derived results indicate that the true speeds (decelerations) of the  EUV waves were about \speed{694, 664, 712, and 648} (\accel{1.175, 1.101, 1.219, and 0.985}), respectively. The initial mean speeds of the first three EUV waves were also derived below a distance of 220 Mm from the eruption source region, and their values are about \speed{907, 962, and 936}, respectively. With the same method, the true initial speeds of the first three EUV waves were \speed{921, 977, and 950}, respectively. One can see that the initial speed of the EUV waves are much faster than those during the entire lifetime of the waves. It is noted that the initial speeds of the waves are similar to those found in \cite{2012ApJ...752L..23S}, where the EUV wave was associated with a powerful X6.9 flare and a halo CME. The derived parameters of the EUV waves are also listed in \tbl{tbl1} for comparison. The measuring results indicate that the observed EUV waves were very fast and showed strong decelerations. In addition, these results also indicate a clear relationship between the wave speeds and the corresponding decelerations, i.e., a faster wave has a stronger deceleration. This suggests that the observed EUV waves were large-amplitude  nonlinear fast-mode magneticsonic waves in nature, in agreement with the results found in  previous studies \citep[e.g.,][]{2011A&A...532A.151W,2014SoPh..289.4563M,2017SoPh..292..185L}.

The kinematics of the four recurrent jets are shown in \nfig{fig5} (b) with the AIA 304 \AA\ time-distance stack plot, in which the first three jets can be identified within a distance of about 60 Mm from the jet-base, and the stripes are shown as straight structures without obvious deceleration. The fourth jet can be clearly observed within a distance more than 400 Mm with an obvious deceleration. It is measured that the ejection speeds of the four jets were about \speed{222, 247, 320, and 342}, respectively. Here, the speed of the fourth jet was obtained by applying a linear fit to the bright stripe below 100 Mm. With the same method as used to derive the true speed of the EUV waves, it is obtained that the true speeds of the jets are about \speed{226, 251, 325, and 347}.

\begin{figure*}
\epsscale{1}
\figurenum{6}
\plotone{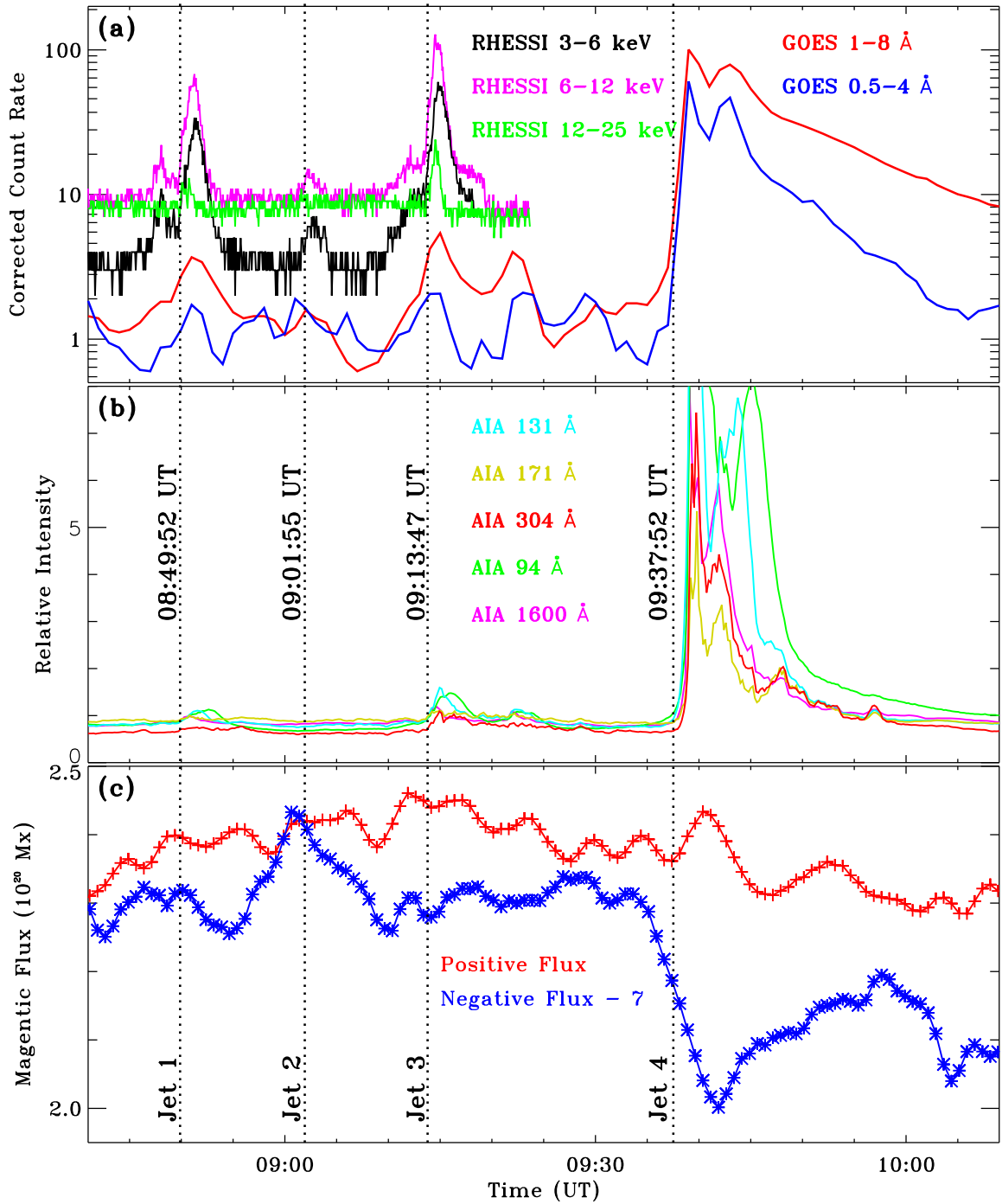}
\caption{Panel (a) shows the X-ray fluxes recorded by the {\em RHESSI} and {\em GOES} instruments, and the different energy band plotted with different colors. Panel (b) shows the AIA lightcurves of the eruption source region (the box region shown in \nfig{fig1} (f)) in different wavelength channels. Panel (c) shows the variations of the positive (red) and negative (blue) magnetic fluxes. The four dotted vertical lines indicate the start times of the four radio type \uppercase\expandafter{\romannumeral3} bursts, respectively.
\label{fig6}}
\end{figure*}

The temporal variations of the {\em RHESSI} and {\em GOES} x-ray fluxes, the EUV relative intensity lightcurves of the flares, and the photospheric magnetic fluxes within the eruption source region are plotted in \nfig{fig6} (a), (b), and (c), respectively. In addition, the start times of the four homologous jets, which are determined based on the start times of the radio type \uppercase\expandafter{\romannumeral3} bursts, are indicated in \nfig{fig6} with four vertical dotted lines. It is clear that all the four jets were formed during the impulsive rising phases of the associated flares. By comparing the start times of the jets and the variations of the magnetic fluxes of the eruption source region (see the white dashed box in \nfig{fig1} (b)), one can find that these jets are tightly associated with the emergence and cancellation events of the photospheric magnetic flux. Right before the start of the first two jets, both the positive and negative fluxes showed obvious increase, then they changed into decrease after the start of the jets. This suggests that the positive and negative fluxes firstly emerged and then cancelled in the eruption source region. The variation trends of the fluxes before and after the start of the third jet were opposite with the first two. For the fourth jet, both the positive and negative fluxes showed obvious decrease right before the jet, then the positive flux started to increase and while the negative flux kept decreasing. This suggests that magnetic cancellation occurred right before the jet, but positive flux started to emerge after the jet. The close temporal relationship between the jets and the flux variations indicate that the occurrence of the observed recurrent jets were tightly in association with the alternating emergence and cancellation magnetic events in the photosphere.

The relevant parameters of the flares, radio type \uppercase\expandafter{\romannumeral3} bursts, coronal jets, and EUV waves are listed in \tbl{tbl1}. Based on the {\em GOES} 1 -- 8 \AA\ soft X-ray flux, the classes of the flares are B4.9, B4.4, B5.2, and C2.5, respectively. It can be seen that the fourth jet was associated with the most energetic C2.5 flare, which may suggest that larger flares can provide more energy to accelerate the jets, like the results found in \cite{2011RAA....11..594S}. The start times of the jets, radio type \uppercase\expandafter{\romannumeral3} bursts, and the waves are determined based on the radio spectrum, AIA 304 \AA\, and AIA 171 \AA\ observations, respectively. By comparing the start times of the flares, radio bursts, and the EUV waves, one can find that the appearance of the radio bursts was delayed from the flare start by 2 -- 5 minutes, while the start times of the jets were delayed with respect to the radio bursts by 1 -- 3 minutes except for the first event. In addition, the appearance of the EUV waves was delayed with respect to the start times of the radio type \uppercase\expandafter{\romannumeral3} bursts and jets by 2 -- 13 and 4 -- 9 minutes, respectively. Besides the second flare, it is interesting that both of the speeds of the jets and the waves are positively related to the flare classes. Namely, more energy released in a flare would drive a faster jet. However, the speeds of the the four EUV waves did not show much difference, although both the associated flares and jets showed relative large energy variation. This may manifest the same background plasma medium in which the waves propagated.

\section{Discussion and Conclusion}
We studied the generation and evolution of four homologous EUV waves originated from active region AR11476 on 2012 May 14 within about one hour from 08:45:00 to 09:45:00 UT, combining the high temporal and high spatial resolution observations taken by {\em SDO} and {\em STEREO}-A. The waves propagated along a transequatorial loop system that connected AR11476 in the northern hemisphere and AR11480 in the southern hemisphere, and the wavefronts were observed as narrow arc-shaped bright features. To our knowledge, the present study is the third detailed analysis of homologous EUV waves in the literature, and new clues are found for diagnosing the driving mechanism and the physical nature of EUV waves. The homologous EUV waves showed obvious deceleration during their propagation; the mean speeds are about \speed{694, 664, 712, and 648}, while their corresponding decelerations are about \accel{1.175, 1.101, 1.219, and 0.985}, respectively. The initial speeds of the first three EUV waves are also derived, and their values are \speed{921, 977, and 950}, respectively. These results indicate that the EUV waves are fast and show strong deceleration; in the meantime, a clear relationship between the wave speeds and the corresponding decelerations can be found, i.e., a faster wave has a stronger deceleration. These properties of the EUV waves suggest that they are large-amplitude nonlinear fast-mode magneticsonic waves or shocks in the physical nature, in agreement with the results found in previous studies \citep[e.g.,][]{2011A&A...532A.151W,2014SoPh..289.4563M,2017SoPh..292..185L}.

We find that the homogenous EUV waves were accompanied by weak flares (B4.4 -- C2.5), coronal jets, and radio type \uppercase\expandafter{\romannumeral3} bursts. Measurements indicate that the speeds of the four jets are about \speed{226, 251, 325, and 347}, respectively. Obviously, the speeds of the coronal jets are much slower than those of the corresponding EUV waves. The EUV waves first appeared ahead of the ejected jets, and the distances to the eruption source region were about 120 Mm. The appearance times of the EUV waves were delayed with respect to the start times of the associated jets about 4 -- 9 minutes. For the driving mechanism of EUV waves, a large number of studies have proposed that EUV waves are possibly driven by flare pressure pulses or CMEs, especially, most of the recent studies based on high resolution data suggested that EUV waves are driven by CMEs rather than flare pressure pulses \citep{2012ApJ...754....7S,2012ApJ...752L..23S,2017ApJ...851..101S,2014SoPh..289.3233L,2015LRSP...12....3W}. For the present case, it is hard to consider that the observed EUV waves were driven by the associated weak flares. On the other hand, it is also hard to consider the scenario that the observed EUV waves were driven by CMEs, since no associated CMEs were detected for the first three waves. The fourth EUV wave was associated with a jet-like CME, however, it just the extension of the jet into the FOVs of the coronagraphs. Therefore, based on the observational results and the close temporal and spatial relationship between the jets and the waves, for the first time, we propose that the observed EUV waves were directly driven by the associated coronal jets.

So far, observations of homologous EUV waves are very scarce in the literature. Some important parameters of the previous two cases (E1: 2010 April 28 \citep{2011ApJ...727L..43K}; E2: 2010 November 11 \citep{2012ApJ...747...67Z}) and the present one (E3: 2012 May 14) are listed in \tbl{tbl2} to compare their similarities and differences. It is interesting that all the three events included four EUV waves traveling in the same direction, and each EUV wave was accompanied by a small flare. E3 has the shortest duration of about one hour, but the speeds of the EUV waves are the fastest among the three events. In addition, the time intervals between two consecutive waves were 105 -- 210, 36 -- 68, and 8 -- 20 minutes in E1, E2, and E3, respectively. The EUV waves in E3 showed obvious deceleration during their propagation, whereas the propagation of the waves in E1 and E2 were almost at constant speeds. The propagation of the EUV waves in E1 and E2 were in quiet-Sun regions, they showed a quasi-circular shape and with a wide angular extent. However, in E3 the propagation of the EUV waves were along a large transequatorial loop system, and they were observed as narrow arc-shaped structures with a small angular extent. The EUV waves in E1 and E2 were all associated with weak flares and faint loop-like CMEs, and also surges in E2. In E3, the waves were associated with weak flares, coronal jets, and radio type \uppercase\expandafter{\romannumeral3} bursts, but without CMEs. 

The parameters and accompanying phenomena of the three homologous EUV wave events indicate that they are very different in physical nature. In previous two cases, loop-like CMEs were observed to be associated with the EUV waves. The quasi-circular shape and wide angular extent of the EUV waves suggests that they were launched by dome-shaped CMEs. It should be pointed out that in E2 the authors observed some surges in the eruption source region and therefore they suggested that the EUV waves were driven by the associated surges. We checked the observations in E2 and found that the surge-like eruptions were the erupting mini-filaments from the source region, and the generation of the EUV waves were not directly driven by the surge-like filament eruptions. In fact, the EUV observations showed that they were excited by the expansion of large-scale coronal loops (i.e., CMEs in the low corona) preceding the erupting filaments, like the generation of the small-scale EUV wave reported by \cite{2017ApJ...851..101S}. Therefore, we believe that the EUV waves in E2 were mostly driven by the associated CMEs. In E3, the EUV waves propagated ahead of the jets along the transequatorial loop system, and no CMEs were associated with the waves. In addition, the narrow arc-shaped wavefronts indicate that the waves were mainly confined and guided by the loop structure. Therefore, it should be reasonable to propose that the observed EUV waves in the present study were directly driven by the associated coronal jets, resemble the generation mechanism of a piston shock in a tube \citep{2015LRSP...12....3W}.

Coronal jets are ubiquitous in the solar atmosphere, and they are often associated with photosphere magnetic flux emergences and cancellations \citep[e.g.,][]{2004ApJ...610.1136L,2005ApJ...628.1056L,2007AA...469..331J,2011ApJ...735L..43S,2012ApJ...745..164S,2012RAA....12..573C,2017ApJ...851...67S,2015ApSS.359...44L}. Previous studies generally proposed that coronal jets are resulting from magnetic reconnection between emerging bipoles and ambient open fields \citep[e.g.,][]{1995Natur.375...42Y,2011ApJ...735L..43S,2011RAA....11.1229Y}. In addition, recent high resolution observations suggest that many jets are dynamically associated with the eruption of mini-filaments \citep[e.g.,][]{2012RAA....12..300Y,2012NewA...17..732Y,2014ApJ...796...73H,2016ApJ...830...60H,2017ApJ...835...35H,2017ApJ...842L..20L,2018ApSS.363...26L}, and often exhibit both cool and hot components \citep[e.g.,][]{2012ApJ...745..164S,2017ApJ...851...67S}. Sometimes, coronal jets play an important role in triggering other large-scale solar eruptions. For example, by interaction with other coronal structures, coronal jets can trigger sympathetic CMEs \citep[e.g.,][]{2008ApJ...677..699J}, filament oscillations \citep[e.g.,][]{2017ApJ...851...47Z}, filament eruptions \citep[e.g.,][]{2016ApJ...827L..12W}, and supply mass to form filaments \citep[e.g.,][]{2005ApJ...631L..93L}. In the present study, it is found that the recurrent jets were tightly associated with alternating flux emergence and cancellations, which are in agreement with previous studies \citep[][; and reference therein]{2016SSRv..201....1R}. In addition, radio type \uppercase\expandafter{\romannumeral3} bursts and {\em RHESSI} X-ray sources are also identified to be in association with the recurrent coronal jets, which indicate the occurrence of magnetic reconnection and acceleration of energetic particles during the ejection processes of the jets. Here, the magnetic morphology as shown in \nfig{fig1} suggested that the jets were possibly produced by the magnetic reconnection between the bright loop structure and the nearby large transequatorial loop system as shown in \nfig{fig1}.

In summary, the present study provide convincing evidence for supporting the scenario that the observed homologous EUV waves were large-amplitude nonlinear fast-mode magnetosonic waves or shocks and they were driven by the associated recurrent coronal jets. This result provides a new driving mechanism for the generation of EUV waves. It is also found that the recurrent jets were tightly in association with the alternating cancellation and emergence of magnetic fluxes in the photosphere, and the radio type \uppercase\expandafter{\romannumeral3} bursts and {\em RHESSI} X-ray sources at the jet-base manifested the magnetic reconnection and acceleration of nonthermal energetic particles during the ejection processes of the jets. More detailed theoretical and statistical studies are need in the further to test the new driving mechanism of EUV waves proposed in the present study.

\acknowledgments
We thank the excellent observations provided by the {\em SDO} and {\em STEREO} teams, and the valuable suggestions of the anonymous referee that largely improved the quality of the present paper. This work is supported by the Natural Science Foundation of China (11773068,11633008, 11403097,11503084, 41774179,11773038), the Yunnan Science Foundation (2015FB191,2017FB006), the Specialized Research Fund for State Key Laboratories, the Open Research Program of CAS Key Laboratory of Solar Activity (KLSA201813), the Youth Innovation Promotion Association (2014047) of Chinese Academy of Sciences, and the grant associated with the Project of the Group for Innovation of Yunnan Province.

%\bibliography{shen}

\begin{thebibliography}{}
\expandafter\ifx\csname natexlab\endcsname\relax\def\natexlab#1{#1}\fi

\bibitem[{{Chen} {et~al.}(2008){Chen}, {Jiang}, \& {Ma}}]{2008AA...478..907C}
{Chen}, H.~D., {Jiang}, Y.~C., \& {Ma}, S.~L. 2008, \aap, 478, 907

\bibitem[{{Chen} {et~al.}(2012){Chen}, {Zhang}, \& {Ma}}]{2012RAA....12..573C}
{Chen}, H.-D., {Zhang}, J., \& {Ma}, S.-L. 2012, Research in Astronomy and
  Astrophysics, 12, 573

\bibitem[{{Chen} {et~al.}(2015){Chen}, {Su}, {Yin}, {Priya}, {Zhang}, {Liu},
  {Xu}, \& {Yu}}]{2015ApJ...815...71C}
{Chen}, J., {Su}, J., {Yin}, Z., {et~al.} 2015, \apj, 815, 71

\bibitem[{{Chen}(2006)}]{2006ApJ...641L.153C}
{Chen}, P.~F. 2006, \apjl, 641, L153

\bibitem[{{Chen} {et~al.}(2016){Chen}, {Fang}, {Chandra}, \&
  {Srivastava}}]{2016SoPh..291.3195C}
{Chen}, P.~F., {Fang}, C., {Chandra}, R., \& {Srivastava}, A.~K. 2016,
  \solphys, 291, 3195

\bibitem[{{Chen} {et~al.}(2002){Chen}, {Wu}, {Shibata}, \&
  {Fang}}]{2002ApJ...572L..99C}
{Chen}, P.~F., {Wu}, S.~T., {Shibata}, K., \& {Fang}, C. 2002, \apjl, 572, L99

\bibitem[{{Cheng} {et~al.}(2012){Cheng}, {Zhang}, {Olmedo}, {Vourlidas},
  {Ding}, \& {Liu}}]{2012ApJ...745L...5C}
{Cheng}, X., {Zhang}, J., {Olmedo}, O., {et~al.} 2012, \apjl, 745, L5

\bibitem[{{Cheung} {et~al.}(2015){Cheung}, {De Pontieu}, {Tarbell}, {Fu},
  {Tian}, {Testa}, {Reeves}, {Mart{\'{\i}}nez-Sykora}, {Boerner}, {W{\"u}lser},
  {Lemen}, {Title}, {Hurlburt}, {Kleint}, {Kankelborg}, {Jaeggli}, {Golub},
  {McKillop}, {Saar}, {Carlsson}, \& {Hansteen}}]{2015ApJ...801...83C}
{Cheung}, M.~C.~M., {De Pontieu}, B., {Tarbell}, T.~D., {et~al.} 2015, \apj,
  801, 83

\bibitem[{{Delaboudini{\`e}re} {et~al.}(1995){Delaboudini{\`e}re}, {Artzner},
  {Brunaud}, {Gabriel}, {Hochedez}, {Millier}, {Song}, {Au}, {Dere}, {Howard},
  {Kreplin}, {Michels}, {Moses}, {Defise}, {Jamar}, {Rochus}, {Chauvineau},
  {Marioge}, {Catura}, {Lemen}, {Shing}, {Stern}, {Gurman}, {Neupert},
  {Maucherat}, {Clette}, {Cugnon}, \& {van Dessel}}]{1995SoPh..162..291D}
{Delaboudini{\`e}re}, J.-P., {Artzner}, G.~E., {Brunaud}, J., {et~al.} 1995,
  \solphys, 162, 291

\bibitem[{{DeVore} \& {Antiochos}(2008)}]{2008ApJ...680..740D}
{DeVore}, C.~R., \& {Antiochos}, S.~K. 2008, \apj, 680, 740

\bibitem[{{Gallagher} \& {Long}(2011)}]{2011SSRv..158..365G}
{Gallagher}, P.~T., \& {Long}, D.~M. 2011, \ssr, 158, 365

\bibitem[{{Gopalswamy} {et~al.}(2009){Gopalswamy}, {Yashiro}, {Temmer},
  {Davila}, {Thompson}, {Jones}, {McAteer}, {Wuelser}, {Freeland}, \&
  {Howard}}]{2009ApJ...691L.123G}
{Gopalswamy}, N., {Yashiro}, S., {Temmer}, M., {et~al.} 2009, \apjl, 691, L123

\bibitem[{{Hong} {et~al.}(2014){Hong}, {Jiang}, {Yang}, {Bi}, {Li}, {Yang}, \&
  {Yang}}]{2014ApJ...796...73H}
{Hong}, J., {Jiang}, Y., {Yang}, J., {et~al.} 2014, \apj, 796, 73

\bibitem[{{Hong} {et~al.}(2017){Hong}, {Jiang}, {Yang}, {Li}, \&
  {Xu}}]{2017ApJ...835...35H}
{Hong}, J., {Jiang}, Y., {Yang}, J., {Li}, H., \& {Xu}, Z. 2017, \apj, 835, 35

\bibitem[{{Hong} {et~al.}(2016){Hong}, {Jiang}, {Yang}, {Yang}, {Xu}, \&
  {Xiang}}]{2016ApJ...830...60H}
{Hong}, J., {Jiang}, Y., {Yang}, J., {et~al.} 2016, \apj, 830, 60

\bibitem[{{Hudson} {et~al.}(2003){Hudson}, {Khan}, {Lemen}, {Nitta}, \&
  {Uchida}}]{2003SoPh..212..121H}
{Hudson}, H.~S., {Khan}, J.~I., {Lemen}, J.~R., {Nitta}, N.~V., \& {Uchida}, Y.
  2003, \solphys, 212, 121

\bibitem[{{Jiang} {et~al.}(2017){Jiang}, {Yan}, {Feng}, {Duan}, {Hu}, {Zuo}, \&
  {Wang}}]{2017ApJ...850....8J}
{Jiang}, C., {Yan}, X., {Feng}, X., {et~al.} 2017, \apj, 850, 8

\bibitem[{{Jiang} {et~al.}(2008){Jiang}, {Shen}, {Yi}, {Yang}, \&
  {Wang}}]{2008ApJ...677..699J}
{Jiang}, Y., {Shen}, Y., {Yi}, B., {Yang}, J., \& {Wang}, J. 2008, \apj, 677,
  699

\bibitem[{{Jiang} {et~al.}(2007){Jiang}, {Chen}, {Li}, {Shen}, \&
  {Yang}}]{2007AA...469..331J}
{Jiang}, Y.~C., {Chen}, H.~D., {Li}, K.~J., {Shen}, Y.~D., \& {Yang}, L.~H.
  2007, \aap, 469, 331

\bibitem[{{Kaiser} {et~al.}(2008){Kaiser}, {Kucera}, {Davila}, {St.~Cyr},
  {Guhathakurta}, \& {Christian}}]{2008SSRv..136....5K}
{Kaiser}, M.~L., {Kucera}, T.~A., {Davila}, J.~M., {et~al.} 2008, \ssr, 136, 5

\bibitem[{{Khan} \& {Aurass}(2002)}]{2002AA...383.1018K}
{Khan}, J.~I., \& {Aurass}, H. 2002, \aap, 383, 1018

\bibitem[{{Kienreich} {et~al.}(2009){Kienreich}, {Temmer}, \&
  {Veronig}}]{2009ApJ...703L.118K}
{Kienreich}, I.~W., {Temmer}, M., \& {Veronig}, A.~M. 2009, \apjl, 703, L118

\bibitem[{{Kienreich} {et~al.}(2011){Kienreich}, {Veronig}, {Muhr}, {Temmer},
  {Vr{\v s}nak}, \& {Nitta}}]{2011ApJ...727L..43K}
{Kienreich}, I.~W., {Veronig}, A.~M., {Muhr}, N., {et~al.} 2011, \apjl, 727,
  L43

\bibitem[{{Krause} {et~al.}(2018){Krause}, {C{\'e}cere}, {Zurbriggen}, {Costa},
  {Francile}, \& {Elaskar}}]{2018MNRAS.474..770K}
{Krause}, G., {C{\'e}cere}, M., {Zurbriggen}, E., {et~al.} 2018, \mnras, 474,
  770

\bibitem[{{Kumar} \& {Manoharan}(2013)}]{2013AA...553A.109K}
{Kumar}, P., \& {Manoharan}, P.~K. 2013, \aap, 553, A109

\bibitem[{{Lemen} {et~al.}(2012){Lemen}, {Title}, {Akin}, {Boerner}, {Chou},
  {Drake}, {Duncan}, {Edwards}, {Friedlaender}, {Heyman}, {Hurlburt}, {Katz},
  {Kushner}, {Levay}, {Lindgren}, {Mathur}, {McFeaters}, {Mitchell}, {Rehse},
  {Schrijver}, {Springer}, {Stern}, {Tarbell}, {Wuelser}, {Wolfson}, {Yanari},
  {Bookbinder}, {Cheimets}, {Caldwell}, {Deluca}, {Gates}, {Golub}, {Park},
  {Podgorski}, {Bush}, {Scherrer}, {Gummin}, {Smith}, {Auker}, {Jerram},
  {Pool}, {Soufli}, {Windt}, {Beardsley}, {Clapp}, {Lang}, \&
  {Waltham}}]{2012SoPh..275...17L}
{Lemen}, J.~R., {Title}, A.~M., {Akin}, D.~J., {et~al.} 2012, \solphys, 275, 17

\bibitem[{{Li} {et~al.}(2018){Li}, {Yang}, {Jiang}, {Bi}, {Qu}, \&
  {Chen}}]{2018ApSS.363...26L}
{Li}, H., {Yang}, J., {Jiang}, Y., {et~al.} 2018, \apss, 363, 26

\bibitem[{{Li} {et~al.}(2017){Li}, {Jiang}, {Yang}, {Qu}, {Yang}, {Xu}, {Bi},
  {Hong}, \& {Chen}}]{2017ApJ...842L..20L}
{Li}, H., {Jiang}, Y., {Yang}, J., {et~al.} 2017, \apjl, 842, L20

\bibitem[{{Li} {et~al.}(2015){Li}, {Jiang}, {Yang}, {Bi}, \&
  {Liang}}]{2015ApSS.359...44L}
{Li}, H.~D., {Jiang}, Y.~C., {Yang}, J.~Y., {Bi}, Y., \& {Liang}, H.~F. 2015,
  \apss, 359, 4

\bibitem[{{Li} {et~al.}(2012){Li}, {Zhang}, {Yang}, \&
  {Liu}}]{2012ApJ...746...13L}
{Li}, T., {Zhang}, J., {Yang}, S., \& {Liu}, W. 2012, \apj, 746, 13

\bibitem[{{Lin} {et~al.}(2002){Lin}, {Dennis}, {Hurford}, {Smith}, {Zehnder},
  {Harvey}, {Curtis}, {Pankow}, {Turin}, {Bester}, {Csillaghy}, {Lewis},
  {Madden}, {van Beek}, {Appleby}, {Raudorf}, {McTiernan}, {Ramaty}, {Schmahl},
  {Schwartz}, {Krucker}, {Abiad}, {Quinn}, {Berg}, {Hashii}, {Sterling},
  {Jackson}, {Pratt}, {Campbell}, {Malone}, {Landis}, {Barrington-Leigh},
  {Slassi-Sennou}, {Cork}, {Clark}, {Amato}, {Orwig}, {Boyle}, {Banks},
  {Shirey}, {Tolbert}, {Zarro}, {Snow}, {Thomsen}, {Henneck}, {McHedlishvili},
  {Ming}, {Fivian}, {Jordan}, {Wanner}, {Crubb}, {Preble}, {Matranga}, {Benz},
  {Hudson}, {Canfield}, {Holman}, {Crannell}, {Kosugi}, {Emslie}, {Vilmer},
  {Brown}, {Johns-Krull}, {Aschwanden}, {Metcalf}, \&
  {Conway}}]{2002SoPh..210....3L}
{Lin}, R.~P., {Dennis}, B.~R., {Hurford}, G.~J., {et~al.} 2002, \solphys, 210,
  3

\bibitem[{{Liu} {et~al.}(2016){Liu}, {Wang}, {Erd{\'e}lyi}, {Liu}, {McIntosh},
  {Gou}, {Chen}, {Liu}, {Liu}, \& {Pan}}]{2016ApJ...833..150L}
{Liu}, J., {Wang}, Y., {Erd{\'e}lyi}, R., {et~al.} 2016, \apj, 833, 150

\bibitem[{{Liu} \& {Ofman}(2014)}]{2014SoPh..289.3233L}
{Liu}, W., \& {Ofman}, L. 2014, \solphys, 289, 3233

\bibitem[{{Liu} {et~al.}(2012){Liu}, {Ofman}, {Nitta}, {Aschwanden},
  {Schrijver}, {Title}, \& {Tarbell}}]{2012ApJ...753...52L}
{Liu}, W., {Ofman}, L., {Nitta}, N.~V., {et~al.} 2012, \apj, 753, 52

\bibitem[{{Liu} {et~al.}(2011{\natexlab{a}}){Liu}, {Title}, {Zhao}, {Ofman},
  {Schrijver}, {Aschwanden}, {De Pontieu}, \& {Tarbell}}]{2011ApJ...736L..13L}
{Liu}, W., {Title}, A.~M., {Zhao}, J., {et~al.} 2011{\natexlab{a}}, \apjl, 736,
  L13

\bibitem[{{Liu} \& {Kurokawa}(2004)}]{2004ApJ...610.1136L}
{Liu}, Y., \& {Kurokawa}, H. 2004, \apj, 610, 1136

\bibitem[{{Liu} {et~al.}(2005{\natexlab{a}}){Liu}, {Kurokawa}, \&
  {Shibata}}]{2005ApJ...631L..93L}
{Liu}, Y., {Kurokawa}, H., \& {Shibata}, K. 2005{\natexlab{a}}, \apjl, 631, L93

\bibitem[{{Liu} {et~al.}(2011{\natexlab{b}}){Liu}, {Luhmann}, {Bale}, \&
  {Lin}}]{2011ApJ...734...84L}
{Liu}, Y., {Luhmann}, J.~G., {Bale}, S.~D., \& {Lin}, R.~P. 2011{\natexlab{b}},
  \apj, 734, 84

\bibitem[{{Liu} {et~al.}(2005{\natexlab{b}}){Liu}, {Su}, {Morimoto},
  {Kurokawa}, \& {Shibata}}]{2005ApJ...628.1056L}
{Liu}, Y., {Su}, J.~T., {Morimoto}, T., {Kurokawa}, H., \& {Shibata}, K.
  2005{\natexlab{b}}, \apj, 628, 1056

\bibitem[{{Liu} {et~al.}(2013){Liu}, {Luhmann}, {Lugaz}, {M{\"o}stl}, {Davies},
  {Bale}, \& {Lin}}]{2013ApJ...769...45L}
{Liu}, Y.~D., {Luhmann}, J.~G., {Lugaz}, N., {et~al.} 2013, \apj, 769, 45

\bibitem[{{Liu} {et~al.}(2014){Liu}, {Richardson}, {Wang}, \&
  {Luhmann}}]{2014ApJ...788L..28L}
{Liu}, Y.~D., {Richardson}, J.~D., {Wang}, C., \& {Luhmann}, J.~G. 2014, \apjl,
  788, L28

\bibitem[{{Long} {et~al.}(2008){Long}, {Gallagher}, {McAteer}, \&
  {Bloomfield}}]{2008ApJ...680L..81L}
{Long}, D.~M., {Gallagher}, P.~T., {McAteer}, R.~T.~J., \& {Bloomfield}, D.~S.
  2008, \apjl, 680, L81

\bibitem[{{Long} {et~al.}(2011){Long}, {Gallagher}, {McAteer}, \&
  {Bloomfield}}]{2011AA...531A..42L}
---. 2011, \aap, 531, A42

\bibitem[{{Long} {et~al.}(2017{\natexlab{a}}){Long}, {Murphy}, {Graham},
  {Carley}, \& {P{\'e}rez-Su{\'a}rez}}]{2017SoPh..292..185L}
{Long}, D.~M., {Murphy}, P., {Graham}, G., {Carley}, E.~P., \&
  {P{\'e}rez-Su{\'a}rez}, D. 2017{\natexlab{a}}, \solphys, 292, 185

\bibitem[{{Long} {et~al.}(2017{\natexlab{b}}){Long}, {Valori},
  {P{\'e}rez-Su{\'a}rez}, {Morton}, \& {V{\'a}squez}}]{2017AA...603A.101L}
{Long}, D.~M., {Valori}, G., {P{\'e}rez-Su{\'a}rez}, D., {Morton}, R.~J., \&
  {V{\'a}squez}, A.~M. 2017{\natexlab{b}}, \aap, 603, A101

\bibitem[{{Long} {et~al.}(2017{\natexlab{c}}){Long}, {Bloomfield}, {Chen},
  {Downs}, {Gallagher}, {Kwon}, {Vanninathan}, {Veronig}, {Vourlidas}, {Vr{\v
  s}nak}, {Warmuth}, \& {{\v Z}ic}}]{2017SoPh..292....7L}
{Long}, D.~M., {Bloomfield}, D.~S., {Chen}, P.~F., {et~al.} 2017{\natexlab{c}},
  \solphys, 292, 7

\bibitem[{{Lugaz} {et~al.}(2017){Lugaz}, {Temmer}, {Wang}, \&
  {Farrugia}}]{2017SoPh..292...64L}
{Lugaz}, N., {Temmer}, M., {Wang}, Y., \& {Farrugia}, C.~J. 2017, \solphys,
  292, 64

\bibitem[{{Luli{\'c}} {et~al.}(2013){Luli{\'c}}, {Vr{\v s}nak}, {{\v Z}ic},
  {Kienreich}, {Muhr}, {Temmer}, \& {Veronig}}]{2013SoPh..286..509L}
{Luli{\'c}}, S., {Vr{\v s}nak}, B., {{\v Z}ic}, T., {et~al.} 2013, \solphys,
  286, 509

\bibitem[{{Ma} {et~al.}(2011){Ma}, {Raymond}, {Golub}, {Lin}, {Chen}, {Grigis},
  {Testa}, \& {Long}}]{2011ApJ...738..160M}
{Ma}, S., {Raymond}, J.~C., {Golub}, L., {et~al.} 2011, \apj, 738, 160

\bibitem[{{Mann} {et~al.}(1999){Mann}, {Aurass}, {Klassen}, {Estel}, \&
  {Thompson}}]{1999ESASP.446..477M}
{Mann}, G., {Aurass}, H., {Klassen}, A., {Estel}, C., \& {Thompson}, B.~J.
  1999, in ESA Special Publication, Vol. 446, 8th SOHO Workshop: Plasma
  Dynamics and Diagnostics in the Solar Transition Region and Corona, ed. J.-C.
  {Vial} \& B.~{Kaldeich-Sch{\"u}}, 477

\bibitem[{{Mei} {et~al.}(2012){Mei}, {Udo}, \& {Lin}}]{2012SCPMA..55.1316M}
{Mei}, Z., {Udo}, Z., \& {Lin}, J. 2012, Science China Physics, Mechanics, and
  Astronomy, 55, 1316

\bibitem[{{Muhr} {et~al.}(2014){Muhr}, {Veronig}, {Kienreich}, {Vr{\v s}nak},
  {Temmer}, \& {Bein}}]{2014SoPh..289.4563M}
{Muhr}, N., {Veronig}, A.~M., {Kienreich}, I.~W., {et~al.} 2014, \solphys, 289,
  4563

\bibitem[{{Narukage} {et~al.}(2008){Narukage}, {Ishii}, {Nagata}, {UeNo},
  {Kitai}, {Kurokawa}, {Akioka}, \& {Shibata}}]{2008ApJ...684L..45N}
{Narukage}, N., {Ishii}, T.~T., {Nagata}, S., {et~al.} 2008, \apjl, 684, L45

\bibitem[{{Nitta} {et~al.}(2014){Nitta}, {Aschwanden}, {Freeland}, {Lemen},
  {W{\"u}lser}, \& {Zarro}}]{2014SoPh..289.1257N}
{Nitta}, N.~V., {Aschwanden}, M.~J., {Freeland}, S.~L., {et~al.} 2014,
  \solphys, 289, 1257

\bibitem[{{Nitta} {et~al.}(2013){Nitta}, {Schrijver}, {Title}, \&
  {Liu}}]{2013ApJ...776...58N}
{Nitta}, N.~V., {Schrijver}, C.~J., {Title}, A.~M., \& {Liu}, W. 2013, \apj,
  776, 58

\bibitem[{{Okamoto} {et~al.}(2004){Okamoto}, {Nakai}, {Keiyama}, {Narukage},
  {UeNo}, {Kitai}, {Kurokawa}, \& {Shibata}}]{2004ApJ...608.1124O}
{Okamoto}, T.~J., {Nakai}, H., {Keiyama}, A., {et~al.} 2004, \apj, 608, 1124

\bibitem[{{Olmedo} {et~al.}(2012){Olmedo}, {Vourlidas}, {Zhang}, \&
  {Cheng}}]{2012ApJ...756..143O}
{Olmedo}, O., {Vourlidas}, A., {Zhang}, J., \& {Cheng}, X. 2012, \apj, 756, 143

\bibitem[{{Panesar} {et~al.}(2016){Panesar}, {Sterling}, \&
  {Moore}}]{2016ApJ...822L..23P}
{Panesar}, N.~K., {Sterling}, A.~C., \& {Moore}, R.~L. 2016, \apjl, 822, L23

\bibitem[{{Pant} {et~al.}(2016){Pant}, {Mazumder}, {Yuan}, {Banerjee},
  {Srivastava}, \& {Shen}}]{2016SoPh..291.3303P}
{Pant}, V., {Mazumder}, R., {Yuan}, D., {et~al.} 2016, \solphys, 291, 3303

\bibitem[{{Pariat} {et~al.}(2010){Pariat}, {Antiochos}, \&
  {DeVore}}]{2010ApJ...714.1762P}
{Pariat}, E., {Antiochos}, S.~K., \& {DeVore}, C.~R. 2010, \apj, 714, 1762

\bibitem[{{Patsourakos} \& {Vourlidas}(2009)}]{2009ApJ...700L.182P}
{Patsourakos}, S., \& {Vourlidas}, A. 2009, \apjl, 700, L182

\bibitem[{{Patsourakos} {et~al.}(2010){Patsourakos}, {Vourlidas}, \&
  {Kliem}}]{2010A&A...522A.100P}
{Patsourakos}, S., {Vourlidas}, A., \& {Kliem}, B. 2010, \aap, 522, A100

\bibitem[{{Patsourakos} {et~al.}(2009){Patsourakos}, {Vourlidas}, {Wang},
  {Stenborg}, \& {Thernisien}}]{2009SoPh..259...49P}
{Patsourakos}, S., {Vourlidas}, A., {Wang}, Y.~M., {Stenborg}, G., \&
  {Thernisien}, A. 2009, \solphys, 259, 49

\bibitem[{{Pesnell} {et~al.}(2012){Pesnell}, {Thompson}, \&
  {Chamberlin}}]{2012SoPh..275....3P}
{Pesnell}, W.~D., {Thompson}, B.~J., \& {Chamberlin}, P.~C. 2012, \solphys,
  275, 3

\bibitem[{{Raouafi} {et~al.}(2016){Raouafi}, {Patsourakos}, {Pariat}, {Young},
  {Sterling}, {Savcheva}, {Shimojo}, {Moreno-Insertis}, {DeVore}, {Archontis},
  {T{\"o}r{\"o}k}, {Mason}, {Curdt}, {Meyer}, {Dalmasse}, \&
  {Matsui}}]{2016SSRv..201....1R}
{Raouafi}, N.~E., {Patsourakos}, S., {Pariat}, E., {et~al.} 2016, \ssr, 201, 1

\bibitem[{{Romano} {et~al.}(2018){Romano}, {Elmhamdi}, {Falco}, {Costa},
  {Kordi}, {Al-Trabulsy}, \& {Al-Shammari}}]{2018ApJ...852L..10R}
{Romano}, P., {Elmhamdi}, A., {Falco}, M., {et~al.} 2018, \apjl, 852, L10

\bibitem[{{Schou} {et~al.}(2012){Schou}, {Borrero}, {Norton}, {Tomczyk},
  {Elmore}, \& {Card}}]{2012SoPh..275..327S}
{Schou}, J., {Borrero}, J.~M., {Norton}, A.~A., {et~al.} 2012, \solphys, 275,
  327

\bibitem[{{Schrijver} \& {De Rosa}(2003)}]{2003SoPh..212..165S}
{Schrijver}, C.~J., \& {De Rosa}, M.~L. 2003, \solphys, 212, 165

\bibitem[{{Shen} {et~al.}(2014{\natexlab{a}}){Shen}, {Ichimoto}, {Ishii},
  {Tian}, {Zhao}, \& {Shibata}}]{2014ApJ...786..151S}
{Shen}, Y., {Ichimoto}, K., {Ishii}, T.~T., {et~al.} 2014{\natexlab{a}}, \apj,
  786, 151

\bibitem[{{Shen} \& {Liu}(2012{\natexlab{a}})}]{2012ApJ...754....7S}
{Shen}, Y., \& {Liu}, Y. 2012{\natexlab{a}}, \apj, 754, 7

\bibitem[{{Shen} \& {Liu}(2012{\natexlab{b}})}]{2012ApJ...753...53S}
---. 2012{\natexlab{b}}, \apj, 753, 53

\bibitem[{{Shen} \& {Liu}(2012{\natexlab{c}})}]{2012ApJ...752L..23S}
---. 2012{\natexlab{c}}, \apjl, 752, L23

\bibitem[{{Shen} {et~al.}(2018{\natexlab{a}}){Shen}, {Liu}, {Song}, \&
  {Tian}}]{2018ApJ...853....1S}
{Shen}, Y., {Liu}, Y., {Song}, T., \& {Tian}, Z. 2018{\natexlab{a}}, \apj, 853,
  1

\bibitem[{{Shen} {et~al.}(2012{\natexlab{a}}){Shen}, {Liu}, \&
  {Su}}]{2012ApJ...750...12S}
{Shen}, Y., {Liu}, Y., \& {Su}, J. 2012{\natexlab{a}}, \apj, 750, 12

\bibitem[{{Shen} {et~al.}(2012{\natexlab{b}}){Shen}, {Liu}, {Su}, \&
  {Deng}}]{2012ApJ...745..164S}
{Shen}, Y., {Liu}, Y., {Su}, J., \& {Deng}, Y. 2012{\natexlab{b}}, \apj, 745,
  164

\bibitem[{{Shen} {et~al.}(2011{\natexlab{a}}){Shen}, {Liu}, {Su}, \&
  {Ibrahim}}]{2011ApJ...735L..43S}
{Shen}, Y., {Liu}, Y., {Su}, J., \& {Ibrahim}, A. 2011{\natexlab{a}}, \apjl,
  735, L43

\bibitem[{{Shen} {et~al.}(2013{\natexlab{a}}){Shen}, {Liu}, {Su}, {Li}, {Zhao},
  {Tian}, {Ichimoto}, \& {Shibata}}]{2013ApJ...773L..33S}
{Shen}, Y., {Liu}, Y., {Su}, J., {et~al.} 2013{\natexlab{a}}, \apjl, 773, L33

\bibitem[{{Shen} {et~al.}(2017{\natexlab{a}}){Shen}, {Liu}, {Tian}, \&
  {Qu}}]{2017ApJ...851..101S}
{Shen}, Y., {Liu}, Y., {Tian}, Z., \& {Qu}, Z. 2017{\natexlab{a}}, \apj, 851,
  101

\bibitem[{{Shen} {et~al.}(2014{\natexlab{b}}){Shen}, {Liu}, {Chen}, \&
  {Ichimoto}}]{2014ApJ...795..130S}
{Shen}, Y., {Liu}, Y.~D., {Chen}, P.~F., \& {Ichimoto}, K. 2014{\natexlab{b}},
  \apj, 795, 130

\bibitem[{{Shen} {et~al.}(2017{\natexlab{b}}){Shen}, {Liu}, {Su}, {Qu}, \&
  {Tian}}]{2017ApJ...851...67S}
{Shen}, Y., {Liu}, Y.~D., {Su}, J., {Qu}, Z., \& {Tian}, Z. 2017{\natexlab{b}},
  \apj, 851, 67

\bibitem[{{Shen} {et~al.}(2018{\natexlab{b}}){Shen}, {Song}, \&
  {Liu}}]{2018MNRAS.477L...6S}
{Shen}, Y., {Song}, T., \& {Liu}, Y. 2018{\natexlab{b}}, \mnras, 477, L6

\bibitem[{{Shen} {et~al.}(2011{\natexlab{b}}){Shen}, {Liu}, \&
  {Liu}}]{2011RAA....11..594S}
{Shen}, Y.-D., {Liu}, Y., \& {Liu}, R. 2011{\natexlab{b}}, Research in
  Astronomy and Astrophysics, 11, 594

\bibitem[{{Shen} {et~al.}(2013{\natexlab{b}}){Shen}, {Liu}, {Su}, {Li},
  {Zhang}, {Tian}, {Zhao}, \& {Elmhamdi}}]{2013SoPh..288..585S}
{Shen}, Y.-D., {Liu}, Y., {Su}, J.-T., {et~al.} 2013{\natexlab{b}}, \solphys,
  288, 585

\bibitem[{{Shibata} {et~al.}(1994){Shibata}, {Nitta}, {Strong}, {Matsumoto},
  {Yokoyama}, {Hirayama}, {Hudson}, \& {Ogawara}}]{1994ApJ...431L..51S}
{Shibata}, K., {Nitta}, N., {Strong}, K.~T., {et~al.} 1994, \apjl, 431, L51

\bibitem[{{Shibata} {et~al.}(2007){Shibata}, {Nakamura}, {Matsumoto}, {Otsuji},
  {Okamoto}, {Nishizuka}, {Kawate}, {Watanabe}, {Nagata}, {UeNo}, {Kitai},
  {Nozawa}, {Tsuneta}, {Suematsu}, {Ichimoto}, {Shimizu}, {Katsukawa},
  {Tarbell}, {Berger}, {Lites}, {Shine}, \& {Title}}]{2007Sci...318.1591S}
{Shibata}, K., {Nakamura}, T., {Matsumoto}, T., {et~al.} 2007, Science, 318,
  1591

\bibitem[{{Sterling} \& {Moore}(2001)}]{2001JGR...10625227S}
{Sterling}, A.~C., \& {Moore}, R.~L. 2001, \jgr, 106, 25227

\bibitem[{{Sui} {et~al.}(2004){Sui}, {Holman}, \&
  {Dennis}}]{2004ApJ...612..546S}
{Sui}, L., {Holman}, G.~D., \& {Dennis}, B.~R. 2004, \apj, 612, 546

\bibitem[{{Thompson} \& {Myers}(2009)}]{2009ApJS..183..225T}
{Thompson}, B.~J., \& {Myers}, D.~C. 2009, \apjs, 183, 225

\bibitem[{{Thompson} {et~al.}(1998){Thompson}, {Plunkett}, {Gurman}, {Newmark},
  {St.~Cyr}, \& {Michels}}]{1998GeoRL..25.2465T}
{Thompson}, B.~J., {Plunkett}, S.~P., {Gurman}, J.~B., {et~al.} 1998, \grl, 25,
  2465

\bibitem[{{Thompson} {et~al.}(1999){Thompson}, {Gurman}, {Neupert}, {Newmark},
  {Delaboudini{\`e}re}, {Cyr}, {Stezelberger}, {Dere}, {Howard}, \&
  {Michels}}]{1999ApJ...517L.151T}
{Thompson}, B.~J., {Gurman}, J.~B., {Neupert}, W.~M., {et~al.} 1999, \apjl,
  517, L151

\bibitem[{{Tian} {et~al.}(2017){Tian}, {Liu}, {Shen}, {Elmhamdi}, {Su}, {Liu},
  \& {Kordi}}]{2017ApJ...845...94T}
{Tian}, Z., {Liu}, Y., {Shen}, Y., {et~al.} 2017, \apj, 845, 94

\bibitem[{{Uchida}(1968)}]{1968SoPh....4...30U}
{Uchida}, Y. 1968, \solphys, 4, 30

\bibitem[{{Vemareddy}(2017)}]{2017ApJ...845...59V}
{Vemareddy}, P. 2017, \apj, 845, 59

\bibitem[{{Vr{\v s}nak} {et~al.}(2005){Vr{\v s}nak}, {Magdaleni{\'c}},
  {Temmer}, {Veronig}, {Warmuth}, {Mann}, {Aurass}, \&
  {Otruba}}]{2005ApJ...625L..67V}
{Vr{\v s}nak}, B., {Magdaleni{\'c}}, J., {Temmer}, M., {et~al.} 2005, \apjl,
  625, L67

\bibitem[{{Vr{\v s}nak} {et~al.}(2016){Vr{\v s}nak}, {{\v Z}ic}, {Luli{\'c}},
  {Temmer}, \& {Veronig}}]{2016SoPh..291...89V}
{Vr{\v s}nak}, B., {{\v Z}ic}, T., {Luli{\'c}}, S., {Temmer}, M., \& {Veronig},
  A.~M. 2016, \solphys, 291, 89

\bibitem[{{Vr{\v s}nak} {et~al.}(2006){Vr{\v s}nak}, {Warmuth}, {Temmer},
  {Veronig}, {Magdaleni{\'c}}, {Hillaris}, \&
  {Karlick{\'y}}}]{2006AA...448..739V}
{Vr{\v s}nak}, B., {Warmuth}, A., {Temmer}, M., {et~al.} 2006, \aap, 448, 739

\bibitem[{{Wang} \& {Liu}(2012)}]{2012ApJ...760..101W}
{Wang}, H., \& {Liu}, C. 2012, \apj, 760, 101

\bibitem[{{Wang} {et~al.}(2009){Wang}, {Shen}, \& {Lin}}]{2009ApJ...700.1716W}
{Wang}, H., {Shen}, C., \& {Lin}, J. 2009, \apj, 700, 1716

\bibitem[{{Wang} {et~al.}(2014){Wang}, {Liu}, {Yang}, \&
  {Hu}}]{2014ApJ...791...84W}
{Wang}, R., {Liu}, Y.~D., {Yang}, Z., \& {Hu}, H. 2014, \apj, 791, 84

\bibitem[{{Wang} {et~al.}(2016){Wang}, {Liu}, {Zimovets}, {Hu}, {Dai}, \&
  {Yang}}]{2016ApJ...827L..12W}
{Wang}, R., {Liu}, Y.~D., {Zimovets}, I., {et~al.} 2016, \apjl, 827, L12

\bibitem[{{Wang} {et~al.}(2013){Wang}, {Liu}, {Shen}, {Liu}, {Ye}, \&
  {Wang}}]{2013ApJ...763L..43W}
{Wang}, Y., {Liu}, L., {Shen}, C., {et~al.} 2013, \apjl, 763, L43

\bibitem[{{Warmuth}(2010)}]{2010AdSpR..45..527W}
{Warmuth}, A. 2010, Advances in Space Research, 45, 527

\bibitem[{{Warmuth}(2015)}]{2015LRSP...12....3W}
---. 2015, Living Reviews in Solar Physics, 12, 3

\bibitem[{{Warmuth} \& {Mann}(2005)}]{2005A&A...435.1123W}
{Warmuth}, A., \& {Mann}, G. 2005, \aap, 435, 1123

\bibitem[{{Warmuth} \& {Mann}(2011)}]{2011A&A...532A.151W}
---. 2011, \aap, 532, A151

\bibitem[{{Warmuth} {et~al.}(2004){Warmuth}, {Vr{\v s}nak}, {Magdaleni{\'c}},
  {Hanslmeier}, \& {Otruba}}]{2004AA...418.1117W}
{Warmuth}, A., {Vr{\v s}nak}, B., {Magdaleni{\'c}}, J., {Hanslmeier}, A., \&
  {Otruba}, W. 2004, \aap, 418, 1117

\bibitem[{{West} {et~al.}(2011){West}, {Zhukov}, {Dolla}, \&
  {Rodriguez}}]{2011ApJ...730..122W}
{West}, M.~J., {Zhukov}, A.~N., {Dolla}, L., \& {Rodriguez}, L. 2011, \apj,
  730, 122

\bibitem[{{Wills-Davey} \& {Thompson}(1999)}]{1999SoPh..190..467W}
{Wills-Davey}, M.~J., \& {Thompson}, B.~J. 1999, \solphys, 190, 467

\bibitem[{{Wuelser} {et~al.}(2004){Wuelser}, {Lemen}, {Tarbell}, {Wolfson},
  {Cannon}, {Carpenter}, {Duncan}, {Gradwohl}, {Meyer}, {Moore}, {Navarro},
  {Pearson}, {Rossi}, {Springer}, {Howard}, {Moses}, {Newmark},
  {Delaboudiniere}, {Artzner}, {Auchere}, {Bougnet}, {Bouyries}, {Bridou},
  {Clotaire}, {Colas}, {Delmotte}, {Jerome}, {Lamare}, {Mercier}, {Mullot},
  {Ravet}, {Song}, {Bothmer}, \& {Deutsch}}]{2004SPIE.5171..111W}
{Wuelser}, J.-P., {Lemen}, J.~R., {Tarbell}, T.~D., {et~al.} 2004, in
  \procspie, Vol. 5171, Telescopes and Instrumentation for Solar Astrophysics,
  ed. S.~{Fineschi} \& M.~A. {Gummin}, 111--122

\bibitem[{{Xu} {et~al.}(2017){Xu}, {Yang}, {Guo}, {Zhao}, {Zhao}, \&
  {Kashapova}}]{2017ApJ...851...30X}
{Xu}, Z., {Yang}, K., {Guo}, Y., {et~al.} 2017, \apj, 851, 30

\bibitem[{{Xue} {et~al.}(2013){Xue}, {Qu}, {Yan}, {Zhao}, \&
  {Ma}}]{2013AA...556A.152X}
{Xue}, Z.~K., {Qu}, Z.~Q., {Yan}, X.~L., {Zhao}, L., \& {Ma}, L. 2013, \aap,
  556, A152

\bibitem[{{Yan} {et~al.}(2012){Yan}, {Qu}, \& {Kong}}]{2012AJ....143...56Y}
{Yan}, X.~L., {Qu}, Z.~Q., \& {Kong}, D.~F. 2012, \aj, 143, 56

\bibitem[{{Yan} {et~al.}(2018){Yan}, {Wang}, {Pan}, {Kong}, {Xue}, {Yang},
  {Li}, \& {Feng}}]{2018ApJ...856...79Y}
{Yan}, X.~L., {Wang}, J.~C., {Pan}, G.~M., {et~al.} 2018, \apj, 856, 79

\bibitem[{{Yang} {et~al.}(2012{\natexlab{a}}){Yang}, {Jiang}, {Yang}, {Hong},
  {Yang}, {Bi}, {Zheng}, \& {Li}}]{2012NewA...17..732Y}
{Yang}, J., {Jiang}, Y., {Yang}, B., {et~al.} 2012{\natexlab{a}}, \na, 17, 732

\bibitem[{{Yang} {et~al.}(2012{\natexlab{b}}){Yang}, {Jiang}, {Zheng}, {Bi},
  {Hong}, \& {Yang}}]{2012ApJ...745....9Y}
{Yang}, J., {Jiang}, Y., {Zheng}, R., {et~al.} 2012{\natexlab{b}}, \apj, 745, 9

\bibitem[{{Yang} {et~al.}(2012{\natexlab{c}}){Yang}, {Jiang}, {Yang}, {Bi},
  {Yang}, {Zheng}, \& {Hong}}]{2012RAA....12..300Y}
{Yang}, J.-Y., {Jiang}, Y.-C., {Yang}, D., {et~al.} 2012{\natexlab{c}},
  Research in Astronomy and Astrophysics, 12, 300

\bibitem[{{Yang} {et~al.}(2013){Yang}, {Zhang}, {Liu}, {Li}, \&
  {Shen}}]{2013ApJ...775...39Y}
{Yang}, L., {Zhang}, J., {Liu}, W., {Li}, T., \& {Shen}, Y. 2013, \apj, 775, 39

\bibitem[{{Yang} {et~al.}(2011){Yang}, {Jiang}, {Yang}, {Bi}, {Zheng}, \&
  {Hong}}]{2011RAA....11.1229Y}
{Yang}, L.-H., {Jiang}, Y.-C., {Yang}, J.-Y., {et~al.} 2011, Research in
  Astronomy and Astrophysics, 11, 1229

\bibitem[{{Yokoyama} \& {Shibata}(1995)}]{1995Natur.375...42Y}
{Yokoyama}, T., \& {Shibata}, K. 1995, \nat, 375, 42

\bibitem[{{Yu} {et~al.}(2014){Yu}, {Zhang}, {Li}, {Zhang}, \&
  {Yang}}]{2014ApJ...782L..15Y}
{Yu}, X., {Zhang}, J., {Li}, T., {Zhang}, Y., \& {Yang}, S. 2014, \apjl, 782,
  L15

\bibitem[{{Zhang} \& {Wang}(2002)}]{2002ApJ...566L.117Z}
{Zhang}, J., \& {Wang}, J. 2002, \apjl, 566, L117

\bibitem[{{Zhang} \& {Ji}(2018)}]{2018arXiv180501088Z}
{Zhang}, Q.~M., \& {Ji}, H.~S. 2018, ArXiv e-prints, arXiv:1805.01088

\bibitem[{{Zhang} {et~al.}(2016){Zhang}, {Ji}, \& {Su}}]{2016SoPh..291..859Z}
{Zhang}, Q.~M., {Ji}, H.~S., \& {Su}, Y.~N. 2016, \solphys, 291, 859

\bibitem[{{Zhang} {et~al.}(2017){Zhang}, {Li}, \& {Ning}}]{2017ApJ...851...47Z}
{Zhang}, Q.~M., {Li}, D., \& {Ning}, Z.~J. 2017, \apj, 851, 47

\bibitem[{{Zhantayev} {et~al.}(2014){Zhantayev}, {Kryakunova}, {Nikolayevskiy},
  \& {Zhumabayev}}]{2014SunGe...9...71Z}
{Zhantayev}, Z., {Kryakunova}, O., {Nikolayevskiy}, N., \& {Zhumabayev}, B.
  2014, Sun and Geosphere, 9, 71

\bibitem[{{Zheng} {et~al.}(2012{\natexlab{a}}){Zheng}, {Jiang}, {Yang}, {Bi},
  {Hong}, {Yang}, \& {Yang}}]{2012AA...541A..49Z}
{Zheng}, R., {Jiang}, Y., {Yang}, J., {et~al.} 2012{\natexlab{a}}, \aap, 541,
  A49

\bibitem[{{Zheng} {et~al.}(2012{\natexlab{b}}){Zheng}, {Jiang}, {Yang}, {Bi},
  {Hong}, {Yang}, \& {Yang}}]{2012ApJ...747...67Z}
---. 2012{\natexlab{b}}, \apj, 747, 67

\bibitem[{{Zheng} {et~al.}(2012{\natexlab{c}}){Zheng}, {Jiang}, {Yang}, {Bi},
  {Hong}, {Yang}, \& {Yang}}]{2012ApJ...753..112Z}
---. 2012{\natexlab{c}}, \apj, 753, 112

\bibitem[{{Zheng} {et~al.}(2012{\natexlab{d}}){Zheng}, {Jiang}, {Yang}, {Bi},
  {Hong}, {Yang}, \& {Yang}}]{2012ApJ...753L..29Z}
---. 2012{\natexlab{d}}, \apjl, 753, L29

\bibitem[{{Zheng} {et~al.}(2013){Zheng}, {Jiang}, {Yang}, {Hong}, {Bi}, {Yang},
  \& {Yang}}]{2013MNRAS.431.1359Z}
{Zheng}, R.-S., {Jiang}, Y.-C., {Yang}, J.-Y., {et~al.} 2013, \mnras, 431, 1359

\bibitem[{{Zong} \& {Dai}(2015)}]{2015ApJ...809..151Z}
{Zong}, W., \& {Dai}, Y. 2015, \apj, 809, 151

\bibitem[{{Zong} \& {Dai}(2017)}]{2017ApJ...834L..15Z}
---. 2017, \apjl, 834, L15

\end{thebibliography}
%\bibliographystyle{aasjournal}

\begin{longrotatetable}
\begin{deluxetable*}{ccccccccccc}
%\tablenum{1}
\tablecaption{Parameters of the three homologous EUV wave events \label{tbl2}}
%\tablewidth{600pt}
\tablehead{
\colhead{Events} & \colhead{Durations} & \colhead{No. Waves} & \colhead{Speeds} & \colhead{Deceleration} &\colhead{Time Interval} &\colhead{Angular} &\colhead{Region} & \multicolumn3c{Accompanying Phenonena }\\
\colhead{yyyy-mm-dd} & \colhead{(hour)} & \colhead{ } & \colhead{(km s$^{-1}$)} & \colhead{(km s$^{-2}$)} &\colhead{(minute)} & \colhead{ } &\colhead{ } & \colhead{ } & \colhead{ } & \colhead{ }
}
%\decimalcolnumbers
\startdata
2010-04-28 &8 &4 &220--340 &$\approx 0$    &105--210 &Wide &Quiet-Sun &Flares and CMEs \\
2010-11-11 &3 &4 &280--500 &\nodata   &36--68 &Wide &Quiet-Sun &Flares, CMEs, and surges \\
2012-05-14 &1 &4 &648--712 &0.985--1.219  & 8--20 & Narrow &Along loop&Flares, jets, and radio type \uppercase\expandafter{\romannumeral3} bursts \\
\enddata
\tablecomments{The events on 2010 April 28 and 2010 November 11 have been documented in \cite{2011ApJ...727L..43K} and \cite{2012ApJ...747...67Z}, respectively. The last event on 2012 May 14 is analyzed in the present study.}
\end{deluxetable*}
\end{longrotatetable}

\end{document}